\documentclass{aa}

\usepackage{graphicx}
\usepackage{txfonts}
\usepackage{upgreek}
\usepackage{placeins}

\begin{document}

   \title{$N$-body simulations of dark matter-baryon interactions}
    \titlerunning{$N$-body simulations of dark matter-baryon interactions}

   \author{Moritz S.\ Fischer
          \inst{\ref{inst:usm},\ref{inst:origins}},
          Klaus Dolag\inst{\ref{inst:usm},\ref{inst:mpa}},
          Mathias Garny\inst{\ref{inst:tum}},
          Vera Gluscevic\inst{\ref{inst:usc}},
          Frederick Groth\inst{\ref{inst:usm}},
          Ethan O.~Nadler\inst{\ref{inst:ucsd}}
          }
    \authorrunning{M.\ S.\ Fischer et al.}

    \institute{
        Universitäts-Sternwarte, Fakultät für Physik, Ludwig-Maximilians-Universität München, Scheinerstr.\ 1, D-81679 München,\\ Germany\label{inst:usm}\\
        \email{mfischer@usm.lmu.de}
        \and
        Excellence Cluster ORIGINS, Boltzmannstrasse 2, D-85748 Garching, Germany\label{inst:origins}
        \and
        Max-Planck-Institut f\"ur Astrophysik, Karl-Schwarzschild-Str. 1, D-85748 Garching, Germany\label{inst:mpa}
        \and
        Physik Department T31, Technische Universit\"at M\"unchen, James-Franck-Straße 1, D-85748 Garching, Germany\label{inst:tum}
        \and
        Department of Physics and Astronomy, University of Southern California, Los Angeles, CA 90089, USA\label{inst:usc}
        \and Department of Astronomy \& Astrophysics, University of California, San Diego, La Jolla, CA 92093, USA\label{inst:ucsd}
    }

   \date{Received 1 April 2025 / Accepted 28 June 2025}

  \abstract
   {Dark matter (DM) particles can interact with particles characterised by the standard model. Although there are a number of constraints derived from direct and indirect detection experiments, the dynamical evolution of astrophysical objects could offer a promising probe for such interactions.
   Obtaining astrophysical predictions is challenging and primarily limited by our ability to simulate scattering between DM and baryonic particles within $N$-body and hydrodynamics simulations.}
   {We have developed the first scheme allowing for the simulation of these interacting dark matter (IDM) models,  accurately accounting for their angular and velocity dependence, as well as the mass ratio between the DM and baryonic scattering partners.}
   {To describe DM-baryon interactions, we used an $N$-body code together with its implementation of smoothed-particle hydrodynamics (SPH) and meshless finite mass.
   The interaction itself was realised in a pairwise fashion by creating a virtual scattering partner from the baryonic particle and allowing it to interact with a DM particle using a scattering routine initially developed for self-interacting dark matter (SIDM). After the interaction, the virtual particle is rejoined with the baryonic particle, fulfilling the requirements of energy and momentum conservation.}
   {Through several test problems, we demonstrated that we are able to reproduce the  analytic solutions with our IDM scheme. This includes a test for scattering with a physical mass ratio of 1:1000, which is beyond the limits of current SIDM simulations.
   We comment on various numerical aspects and challenges, and we describe the limitations of our numerical scheme. Furthermore, we study the impact of IDM on halo formation with a collapsing over-density.}
   {We find that it is possible to accurately model IDM within $N$-body and hydrodynamics simulations commonly used in astrophysics. Finally, our scheme allows for  novel predictions to be made and new constraints on DM-baryon scattering to be set.}


   \maketitle

\section{Introduction}

Despite various efforts to decipher the nature of dark matter (DM) with laboratory experiments, all evidence for its existence still stems from astrophysical and cosmological observations, such as the rotation curves of galaxies \citep{Rubin_1980, Bosma_1981}. Even though the collisionless cold dark matter (CDM) model is quite successful in explaining several observations, including the cosmological large-scale structure, it does not provide much insight into the particle nature of DM. Whether DM has any interaction other than the gravitational force remains an open question. Potentially, DM particles could interact with each other through novel physics of a dark sector or end up coupled to the particles of the standard model (SM).

Both scenarios have the potential to change the evolution of astrophysical objects such as galaxies and galaxy clusters, compared to the case of collisionless DM in cases where the interactions would be strong enough.
This opens up a window onto probing the particle nature of DM via astronomy and eventually discovering new physics of the dark sector. It also paves an avenue for probing the cross-section for DM self-interactions and DM-baryon interactions.

In the first case of self-interacting dark matter (SIDM), various astrophysical systems and their observables have been used to constrain the strength of self-interactions and explain potential discrepancies between CDM predictions and observations \citep[described in the review articles by][]{Tulin_2018, Adhikari_2022}. The strongest upper bounds on the cross-section come from galaxy clusters, while low-mass systems such as dwarf galaxies offer hints that DM may have strong self-interactions at low velocities. These efforts have been supported by semi-analytical and numerical modelling of SIDM. In particular, for the latter, it is possible to run full physics cosmological simulations to obtain predictions for the formation of objects covering a large mass range from dwarf galaxies \citep[e.g.][]{Vogelsberger_2014, Correa_2025} through galaxy clusters \citep[e.g.][]{Robertson_2019, Ragagnin_2024, Despali_2025}.

In contrast, for the case of interacting dark matter (IDM), there are no  cosmological simulations available to model the DM-baryon interactions in situ via the simulation codes. The astrophysical constraints are based on analytical calculations and simulations of the linear evolution of the matter power spectrum \citep[e.g.][]{Sigurdson_2004, Dvorkin_2014, Boddy_2018a, Ali-Haimoud_2023}.
They have, for example, been derived from cosmic microwave background (CMB) observations \citep{Boddy_2018b, Gluscevic_2018}, the Lyman-alpha forest and large scale structure \citep{Dvorkin_2014, He_2023, He_2025}, and the Milky Way satellite galaxy abundance \citep{Maamari_2021, Crumrine_2025}.
A recent study \citep{He_2025} constrained velocity-dependent DM-baryon interactions employing CMB data together with observations of the large scale structure.
Constraints on DM-proton and DM-electron scattering were also derived by~\cite{Nguyen_2021} and~\cite{Buen_Abad_2022} based on CMB anisotropies, baryon acoustic oscillations, the Lyman-$\alpha$ forest, and the abundance of Milky Way subhalos. The implications at low redshifts due to an altered matter power spectrum by DM-baryon scattering at high redshifts were recently studied using $N$-body simulations, assuming collisionless DM \citep[e.g.][]{Zhang_2024, Nadler_2025a, Nadler_2025b, An_2025}.
Moreover, galaxy clusters provide a promising probe for significant DM-baryon interactions at late times. By studying the heat exchange between the DM and the intra-cluster medium, constraints on  IDM can be inferred~\cite[e.g.][]{Shoji_2024, Stuart_2024}.
Furthermore, the ionisation of molecular clouds has been used to constrain DM-proton scattering \citep{Prabhu_2023, Blanco_2023} and the orbital decay of pulsars has been employed in constraining IDM as well \citep{Lucero_2024}.

More generally speaking,  signatures of DM-baryon scattering at low redshift may have similarities to DM self-interactions, since IDM is also expected to alter the matter distribution on small scales. While SIDM is only capable of altering the properties of the baryons indirectly via changes in the DM gravitational potential, the effects of IDM go beyond that. In particular, DM-baryon interactions can lead to energy exchange between DM and baryons, which means that they could cool or heat the interstellar or intra-cluster medium, depending on the particle physics model. Moreover, these interactions can make relative motions between DM and baryons decay or affect the ionisation fraction of the gas. In turn, these effects can potentially impact other processes, such as star formation, and alter the evolution of galaxies. Overall, the phenomenology of IDM is rich and can vary substantially between models with interaction cross-sections that differ in their angular and velocity dependence as well as the mass ratio of the interacting particle species.

The extent to which astrophysical probes can be used to constrain DM-baryon interactions is limited by our ability to model these interactions. The use of $N$-body simulations is a common technique for studying various astrophysical systems at different scales, for example, the formation and evolution of galaxies within the cosmological context. The evolution of galaxies is shaped by various physical processes beyond gravity, such as gas dynamics, the evolution of the stellar component, and black holes (BH). The non-linear interplay of different physical processes often hinders a precise analytic description and requires numerical simulations.  Unfortunately, however, it has not yet been possible to successfully include the effects of DM-baryon interactions in such simulations and thus exploit the full potential of astrophysical probes to learn about those interactions.

On the other hand, significant efforts have been undertaken to constrain DM interactions with direct detection experiments \citep[see for example][and the references therein]{Cirelli_2024}. In particular, ground-based experiments have mainly been carried out, but  studies are not limited to these \citep[e.g.][]{Emken_2019, Du_2024}.
In addition, DM interactions are also tested by searching for annihilation and decay products in cosmic and gamma rays.
This implies a wealth of constraints that could be complemented by studies of the dynamical impact of DM physics.

While numerous constraints on DM-baryon interactions exist, leveraging astrophysical probes with the help of new techniques to model these interactions can provide novel constraints that may allow us to probe uncharted parts of the DM parameter space.
Towards this end, we introduce a novel numerical scheme that allows us to simulate the interaction between SM and DM particles in $N$-body simulations in situ. 
We study its numerical behaviour, show its abilities, and apply it to the collapse of an overdensity and halo formation.
Our aim is to develop a scheme that is suitable for application in cosmological simulations of galaxy formation involving various physical processes. However, the exploration of this possibility lies beyond the scope of this paper. Instead, we focus on the numerical foundation to describe the DM interactions.

This work is structured as follows.
In Sect.~\ref{sec:numerical_methods} we explain the scheme for the DM-baryon interactions. It follows a set of test problems to study its numerical properties (Sect.~\ref{sec:test_problems}).
We study the effect of IDM on the halo formation by simulating the collapse of an overdensity (Sect.~\ref{sec:halo_formation}).
In Sect.~\ref{sec:discussion}, we discuss the limitations of this work as well as directions for further research.
Finally, we present our summary and conclusions in Sect.~\ref{sec:conclusion}.
Additional information is provided in the appendices.

\section{Numerical methods} \label{sec:numerical_methods}

\begin{figure}
    \centering
    \includegraphics[width=\columnwidth]{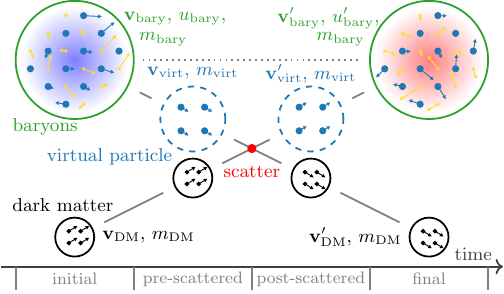}
    \caption{Illustration of the numerical scheme for the DM-baryon interactions. The numerical particles are shown together with the physical particles they represent. The velocities of the physical particles are indicated by small arrows. The different stages in treating the interaction between baryons and DM for a single pair of numerical particles are illustrated from the left to the right. As shown here, the  case of an interaction between a numerical baryonic particle and a numerical DM particle involves a change in their bulk motion and heating of the baryons. Finally, we note that here we also illustrate the case where the baryons consist of two species, but only one interacts with DM.}
    \label{fig:sketch_scheme}
\end{figure}

In this section, we introduce our novel formulation of the interactions between DM and baryons.
A sketch of the idea can be found in Fig.~\ref{fig:sketch_scheme}.
We explain the numerical scheme and describe the implementation.

\subsection{Formulation of DM-baryon interactions} \label{sec:formulation_scheme}
We aim to describe the DM-baryon interactions within $N$-body simulations where the baryons are modelled with smoothed particle hydrodynamics (SPH), meshless finite mass (MFM), or other schemes.
Those codes are commonly used for cosmological simulations and studies of galaxy formation and evolution.

Overall, DM is represented by numerical particles characterised by the mass, $m_\mathrm{DM}$, position, $\mathbf{x}_\mathrm{DM}$, and velocity, $\mathbf{v}_\mathrm{DM}$.
This can be understood as a particle representing a phase-space patch with the physical DM particles having the same velocity.
The numerical particles representing the baryons analogously have a mass, $m_\mathrm{bary}$, and position, $\mathbf{x}_\mathrm{bary}$.
In addition, the bulk velocity of the physical particles is described by $\mathbf{v}_\mathrm{bary}$ and their random motions, assumed to follow a Maxwell-Boltzmann distribution, are characterised by the internal energy per mass, $u_\mathrm{bary}$.

Similarly to the numerical schemes of SIDM, we formulated the interaction based on pairs of numerical particles.
A pair consists always of one DM and one baryonic particle.
The interaction only takes place if they are close enough to each other.
The particles are assigned a kernel with a size, $h$, determined by the $N_\mathrm{ngb}$ next neighbours of the same particle specified (i.e.\ DM or baryons).\footnote{We note that in practice, we define the kernel size used for IDM differently, but it is based on the $h$ parameter described here. Further information can be found in Sect.~\ref{sec:rescaling_kernel_size}.}
They only interact if their kernels are overlapping.
During a time step, for a particle, all the pairs it forms with its neighbours from the other particle species (DM or baryons) are used to model the DM-baryon interactions.
In the following, we describe the details of this interaction.

The physical particles represented by a numerical DM-baryon pair would scatter in different centre-of-mass systems and their post-scattering velocities would point in various directions.
It is impossible to accommodate these distributions with the two numerical particles of the pair.
Instead, we use a stochastic description and develop a Monte Carlo scheme, as commonly done in SIDM as well.
However, in contrast to the DM particle, the baryonic one represents a whole distribution of velocities and not a single one.
This is because for the baryons we can assume that the velocities locally follow a Maxwell-Boltzmann distribution, whereas the DM particles do not follow a specific velocity distribution, making it necessary to resolve the velocity space.\footnote{Given that the $N$-body representation allows for arbitrary velocity distributions to be included, we can describe the effect of the interactions on the DM velocity distribution; for example, turning a cold DM distribution into a warm one or vice versa.}
To account for this, we take a random velocity from the distribution of baryonic velocities.
This velocity is used to form an interaction partner for the numerical DM particle.

Specifically, we create a particle called a `virtual' particle when we want to compute the interaction between a DM-baryon pair and we destroy the virtual particle when we completed the interaction of this pair (as illustrated in Fig.~\ref{fig:sketch_scheme}).
This implies that we create many virtual particles from each considered baryonic particle per time step, but they never exist simultaneously.
This is because all pairs that a baryonic particle forms are computed in a well-defined consecutive manner and a virtual particle exists only for the time we are considering a specific pair of a numerical baryonic and DM particle.
This consecutive order is necessary to ensure energy conservation. If, in contrast, we were to execute the computations of two pairs that share a common particle at the same time, we would use the same initial properties for the computations. Thus, we might end up with two incompatible sets for the post-scattered properties. Despite being consecutive for every particle, the specific order in which the pairwise computations are executed does not matter and, thus, it should not impact the accuracy of our simulations.
The only purpose of the virtual particles is to allow us to formulate the interaction between the DM and baryons in an manner that explicitly conserves the mass, energy, and momentum\footnote{As for SIDM, an explicit conservation of angular momentum is not guaranteed.}.

The virtual particle sits at the same position as the baryonic one, $\mathbf{x}_\mathrm{virt} = \mathbf{x}_\mathrm{bary}$, and it has the same kernel size, $h_\mathrm{virt} = h_\mathrm{bary}$.
Its velocity is given by the bulk motion of the baryons plus a random component, expressed as
\begin{equation}
    \mathbf{v}_\mathrm{virt} = \mathbf{v}_\mathrm{bary} + \mathbf{v}_\mathrm{rand}(u_\mathrm{bary}) \,.
\end{equation}
We note that the random component depends on the internal energy per mass of the baryonic particle, $u_\mathrm{bary}$.
It is drawn from a Maxwell-Boltzmann distribution,
\begin{equation} \label{eq:maxwell}
    f(v) = \sqrt{\frac{2}{\uppi}} \, \frac{v^2}{a^3} \, e^{-\frac{v^2}{2 a^2}}\quad \textnormal{with} \quad a = \sqrt{\frac{2}{3} \, u_\mathrm{bary}}\,.
\end{equation}
To avoid very high velocities for the virtual particle, we cut the high-velocity end of the Maxwell-Boltzmann distribution.
Velocities higher than $v_\mathrm{cut} = \zeta \, a$ are reduced to $v_\mathrm{cut}$.
In practice, we used $\zeta = 5$, which should be large enough such that its effect is negligible compared to all scatterings.
More precisely, $\zeta = 5$, implies a relative error for the energy represented by the velocity distribution of $\approx 10^{-5}$. However, this $v^2$ weight may underestimate the impact on the simulation results. A better estimate might be to weight the velocities by $v^5$, as motivated by the effective or characteristic cross-sections for SIDM \citep{Yang_2022D, Yang_2023S}. With such a weight, the relative error becomes $\approx 3 \times 10^{-4}$. This implicitly assumes a velocity-independent cross-section. However, if the cross-section decreases with velocity, the error would be smaller.

It should be noted that with Eq.~\eqref{eq:maxwell}, we assume that the baryons consist only of one type of particle, namely, the one that interacts with the DM.
However, we could also accommodate for more complicated situations, which may require a different value for the distribution parameter, $a$.

To model the scattering kinematics correctly, the numerical particles must have the same mass ratio, $r$, as the physical particles.\footnote{This can be illustrated with a two-species model consisting of a heavier and a lighter species. In the equilibrium state, the energy equipartition depends on the mass ratio of the particles or in other words the lighter particles exhibit a larger velocity dispersion than the heavier ones. Since the interactions (modelled analogously to particle scattering) conserve energy and linear momentum explicitly, the energy equipartition in the $N$-body system depends on the numerical particle masses in the same way as the physical system depends on the physical particle masses. Therefore, the numerical system accurately represents the physical system only if the numerical mass ratio matches the physical mass ratio.}
For the mass of the virtual particle,
we obtain\begin{equation}
    m_\mathrm{virt} = r \, m_\mathrm{DM} \,.
\end{equation}

Based on the DM and virtual particle, we can compute the scattering as done in SIDM codes \citep[e.g.][]{Koda_2011, Fry_2015, Robertson_2017a, Yang_2022D}.
In practice, we follow \cite{Fischer_2021a}. 
The scattering can alter the velocities of the two numerical particles.
Hence we obtain the post-scattered velocity, $\mathbf{v}'_\mathrm{DM}$, for the DM particle and $\mathbf{v}'_\mathrm{virt}$ for the virtual particle.
The post-scattered velocities are obtained by rotating the momentum vectors in the centre-of-mass frame.
In the case of large-angle scattering, we first compute a probability of determining whether the two particles interact or not \citep[e.g.][]{Burkert_2000, Rocha_2013}.
In particular, we follow the scheme for rare interactions of \cite{Fischer_2021a}.
The interaction strength for the particles $i$ and $j$, depends on a geometrical factor, $\Lambda_{ij}$, based on the kernels, $W(\mathbf{x}, h),$ assigned to the numerical particles, as
\begin{equation} \label{eq:overlap}
    \Lambda_{ij} = \int W(|\mathbf{x}-\mathbf{x}_i|, h_i) \, W(|\mathbf{x}-\mathbf{x}_j|, h_j) \, \mathrm{d}^3\mathbf{x} \,.
\end{equation}
We compute $\Lambda_{ij}$ as described in Appendix~A by \cite{Fischer_2021a}.

The probability that the DM particle and the virtual particle interact is given as
\begin{equation} \label{eq:probability}
    P_{ij} = \frac{\sigma(v)}{m_\chi} \, \frac{f_\mathrm{bary}}{\mu} \, m_\mathrm{DM} \, v \, \Delta t \, \Lambda_{ij} \quad \textnormal{with} \quad v = |\mathbf{v}_i - \mathbf{v}_j| \,.
\end{equation}
The total cross-section is given by $\sigma(v)$, and the physical DM particle mass by $m_\chi$.
We employ the mass ratio $\mu = m_\mathrm{virt} / m_\mathrm{bary}$.
We note that here we use the velocity of the virtual particle, not the baryonic one.
This is a consequence of the fact that the baryonic particle represent a distribution of velocities and the velocity of the virtual particle is a random velocity drawn from this distribution. Using the velocity from the virtual particle gives the correct interaction probability and allows us to account for arbitrary velocity dependencies of the interaction cross-section.
In addition, we introduced the parameter $f_\mathrm{bary}$ to specify the mass fraction of the baryonic particle taking part in the interaction.
It is worth mentioning that here we assume a physical particle scatters only once with a particle represented by the other numerical particle of the pair per time step.
This still allows for a particle to scatter multiple times, but only with partners from different pairs.
The probability that a physical particle scatters twice is $P_{ij}^2$.
Hence, we have to choose the time step, $\Delta t$, that would be small enough to make $P_{ij}^2$ negligible.
This constraint is a consequence of modelling the interactions between two numerical particles analogously to a single physical scattering event; namely we produce the post-scattered velocity distribution by assuming a single scattering event.
For multiple scattering events per particle, this distribution would look differently and, thus, the contribution of those must be kept small to accurately model the interactions.

Similarly to the interaction probability, we can formulate a drag force term for small-angle scattering analogously to the scheme for frequent self-interactions by \cite{Fischer_2021a}.
To characterise the strength of the interaction, we use the momentum transfer cross-section,
\begin{equation} \label{eq:transfer_cross-section}
    \sigma_\mathrm{T} = 2 \uppi \int^1_{-1} \frac{\mathrm{d} \sigma}{\mathrm{d}\Omega_\mathrm{cms}} (1-\cos\theta_\mathrm{cms}) \, \mathrm{d}\cos\theta_\mathrm{cms} \,.
\end{equation}
The drag force for our DM-baryon interactions is given as\begin{equation} \label{eq:drag_force}
    F_\mathrm{drag} = \frac{\sigma_\mathrm{T}(v)}{m_\chi} \, \frac{f_\mathrm{bary}}{\mu \, (1+r)} \, m_\mathrm{virt} \, m_\mathrm{DM} \, v^2 \, \Lambda_{ij} \quad \textnormal{with} \quad v = |\mathbf{v}_i - \mathbf{v}_j| \,.
\end{equation}
A derivation of the interaction probability, $P_{ij}$ (Eq.~\eqref{eq:probability}), and the drag force, $F_\mathrm{drag}$ (Eq.~\eqref{eq:drag_force}), can be found in Appendix~\ref{sec:derivation_p_and_drag}. The final step of the interaction is to destroy the virtual particle or in other words, thermalise it back into the baryonic particle.
The underlying idea is that the interactions between the physical baryonic particles are strong enough to quickly distribute the exchanged momentum and energy from the DM-baryon scattering over the physical baryonic particles (represented by the single numerical particle under consideration).
However, this thermalisation timescale must be small enough (relative to the numerical time step, which is roughly speaking set by the minimum of the local dynamical time and the inverse scattering rate), so that our previous assumption of a Maxwell-Boltzmann distribution is valid as well.

Next, we can derive the post-scattered properties of the baryonic particle.
To do so, we start with momentum conservation to obtain the new bulk motion; namely,\ the momentum change that the virtual particle has experienced is then applied to the baryonic particle.
\begin{equation}
    \mathbf{v}'_\mathrm{bary} = \mathbf{v}_\mathrm{bary} + \mu \, \left( \mathbf{v}'_\mathrm{virt} - \mathbf{v}_\mathrm{virt} \right) \, .
\end{equation}
We also update the internal energy per mass fulfilling energy conservation, namely,\ the baryonic particle must account for energy change that the virtual particle experiences.
\begin{equation}
    u'_\mathrm{bary} = u_\mathrm{bary} + \frac{1}{2} \, \left[ \mathbf{v}^2_\mathrm{bary} - \mathbf{v}'^2_\mathrm{bary} + \frac{m_\mathrm{DM}}{m_\mathrm{bary}} \left( \mathbf{v}^2_\mathrm{DM} - \mathbf{v}'^2_\mathrm{DM} \right)\right] \, .
\end{equation}
This enables the internal energy of the baryons to increase or decrease, effectively allowing for heat flow between the two components in each direction.

We expect the numerical scheme we present here to converge to the true physical solution in the simultaneous limit of $N \rightarrow \infty$, $N_\mathrm{ngb} \rightarrow \infty$, $N / N_\mathrm{ngb} \rightarrow \infty$, $m_\mathrm{bary} / m_\mathrm{DM} \rightarrow \infty$ and $\Delta t \rightarrow 0$.\footnote{When the kernel rescaling technique described in Sect.~\ref{sec:rescaling_kernel_size} is used $N_\mathrm{ngb}$ should be replaced by $N_\mathrm{idm}$.}
We note that for very anisotropic cross-sections in the frequent limit, we no longer require $m_\mathrm{bary} / m_\mathrm{DM} \rightarrow \infty$. This limit is relevant to reduce the effect a single numerical particle interaction has on the baryonic particle, but in the frequent interaction scheme for small-angle scattering, this is already achieved by $\Delta t \rightarrow 0$ or $N_\mathrm{ngb} \rightarrow \infty$.
In the frequent interaction scheme, an effective scattering angle determined by a drag force description is used \citep{Fischer_2021a}.
It also depends on the size of the time step and the neighbour number.
In this case, the effect on the baryonic particle per numerical particle interaction can simply be reduced by choosing smaller time steps or increasing the neighbour number.
This implies, for the limit of very anisotropic cross-sections, that we do not require the numerical mass of the baryonic particle to be much higher than the one of the DM particle.

\subsection{Negative internal energy problem} \label{sec:negative_energy_problem}
The energy required to generate the numerical virtual particle from the baryonic one can potentially be greater than the internal energy of the baryonic particle.
Precisely speaking, in the rest frame of the baryonic particle before creating the virtual particle, the sum of the kinetic energy of the virtual particle and the remaining baryonic one (minus the mass of the virtual particle) could be greater than its initial internal energy. This can cause a problem as in the case when the following inequality is violated,
\begin{equation}
    \underbrace{u_\mathrm{bary} \, m_\mathrm{bary}}_{= E_\mathrm{internal, bary}} > \underbrace{\frac{1}{2} \, m_\mathrm{virt} \, \mathbf{v}^2_\mathrm{virt}}_{=E_\mathrm{kin, virt}} + \underbrace{\frac{1}{2} \frac{m_\mathrm{virt}}{m_\mathrm{bary} - m_\mathrm{virt}} \mathbf{v}^2_\mathrm{virt}}_{=E_\mathrm{kin, rest}} \,.
\end{equation}
If this condition is not fulfilled, the virtual particle could lose enough energy via scattering with a DM particle to the extent that even after the virtual and baryonic particle are rejoined, the internal energy of the baryonic particle is non-positive. This would cause a severe problem for the numerical scheme to model the hydrodynamics.

To ensure that the specific internal energy, $u,$ does not become zero or negative, we find that the velocity of the virtual particle must fulfil
\begin{equation} \label{eq:no_negative_u}
    |\mathbf{v}_\mathrm{rand}| < a \, \sqrt{3 \left(\frac{m_\mathrm{bary}}{r\,m_\mathrm{DM}}-1\right)} = a \, \sqrt{3 \left(\frac{m_\mathrm{bary}}{m_\mathrm{virt}}-1\right)} \, .
\end{equation}
This can be seen by requiring that the remaining internal energy of the baryonic particle must be positive after creating the virtual particle.

When replacing the left-hand side of Eq.~\eqref{eq:no_negative_u} with $v_\mathrm{cut} = a \, \zeta$, because $|\mathbf{v}_\mathrm{rand}|$ is chosen not to be larger than $v_\mathrm{cut}$, we can derive a bound on the mass ratio of the baryonic and DM particles,
\begin{equation} \label{eq:mass_ratio_constraint}
    \frac{m_\mathrm{bary}}{m_\mathrm{DM}} > r \, \left( \frac{\zeta^2}{3} + 1 \right) \,.
\end{equation}
This constrains the mass ratio of the numerical particles required to prevent negative internal energies.
We have to note that Eq.~\eqref{eq:mass_ratio_constraint} only applies to large-angle scattering as the maximal change in internal energy that we use here only arises from a large scattering angle.

In contrast, small scattering angles typically do not lead to enough energy exchange for creating negative values of the internal energy.
But if the numerical particle mass of the baryons is much smaller than the one of the DM particles even small-angle scattering could run into the problem of negative internal energies given that the time step is not chosen small enough.\footnote{We note that reducing the time step may not always solve the problem of being able to advance the simulation in time, but instead heat conduction can play an important role, as explained in Sect.~\ref{sec:viscosity_and_heat_conduction}.} However, this also depends on the baryon temperature and DM velocity dispersion involved. Also,\ there is no equation as simple as Eq.~\eqref{eq:mass_ratio_constraint} for such a case.

For some models, Eq.~\eqref{eq:mass_ratio_constraint} might be to restrictive.
Those featuring large-angle scattering and having a sizeable value of $r$ would require numerical particles for the DM being much less massive than the numerical baryonic particles. This makes simulating models with a large $r$, fairly expensive or even unfeasible. However, at the same time, scatterings in such a model are much less likely to lead to negative internal energy as they have a smaller impact on the baryonic scattering partner. Consequently, it might be more useful to choose a set-up that does not satisfy Eq.~\eqref{eq:mass_ratio_constraint}. To ensure at the same time that internal energies stay positive, we explicitly check the internal energy after each interaction and reject the post-scattered values if the internal energy has a non-positive value. We recompute the scattering event from the pre-scattered values (for large-angle scattering this includes the decision if the particles interact) and repeat this until the internal energy is positive. In practice, for large-angle scattering, this turns most of the problematic scattering events into non-scatters. If too many scatterings are rejected this would lead to an underestimate of the effect of DM-baryon interactions. We note that the scatterings with a high relative velocity are predominately rejected, namely, the ones that lead to the largest energy and momentum exchange between DM and baryons.

We note that when the simulation parameters are chosen carefully, this rejection scheme has a negligible impact on the simulation results, as only a small number of scatterings end up being rejected. With this in mind, we set up the simulations for this work to ensure that rejections either do not occur at all, which applies for most of our simulations, or occur very rarely. In the latter case, the number of rejections is many orders of magnitude smaller than the total number of interactions. In principle, the number of rejections depends on the choice of $\zeta$; for a larger value, more rejections could occur.

\subsection{The role of viscosity and heat conduction} \label{sec:viscosity_and_heat_conduction}
To model DM-baryon interactions accurately with the scheme described above, viscosity plays a crucial role.
This can be illustrated by considering a situation where the DM and baryon distributions are at rest, but have a different velocity dispersion (temperature)\footnote{We note that the temperature for the baryons is represented differently compared to DM. While for DM the velocity space is resolved, it is approximated with a Maxwell--Boltzmann distribution plus bulk motion for the baryons. The internal energy or temperature of the SPH/MFM particles refers to the Maxwell--Boltzmann distribution and their velocity with respect to their bulk motion.
In contrast, we do not assign any internal energy to the DM particles, but only a velocity.
For DM, bulk motion can be defined as the centre of mass velocity of an ensemble of numerical particles, while the motion relative to the centre of mass motion gives rise to a local velocity dispersion, which depending on the velocity distribution could be interpreted as temperature. We also note, that the temperature does not only depend on the velocity dispersion but also on the mass of the physical particles.}.
Physically speaking, the interactions would cool the DM (reduce their velocity dispersion) and heat the baryons or vice versa, while no net momentum would be exchanged and both components would remain at rest. However, in our numerical scheme, we also modify the bulk motion, i.e.\ the velocity of the numerical baryonic particles. This bulk motion is random and not coherent among the numerical particles. It might be viewed as an artificial small-scale turbulence. The strength of this turbulence could potentially grow over the course of the simulation. How strong it is depends on the numerical set-up, for example, the value of $\mu$.

To ensure accurate results, the artificial turbulent motion of the baryons (random bulk motion) must be kept small. If it reaches a significant strength though, the temperature of the baryons would appear to be cooler compared to the exact physical solution. If the baryons are subject to viscous forces, turbulent motion, including artificial turbulence, is dampened. Viscosity effectively transfers the kinetic energy induced by the DM-baryon interactions into the internal energy of the baryons and thus increases the temperature of the baryons. Overall, viscosity must be strong enough to ensure that the heat exchanged between baryons and DM does not significantly end up as kinetic energy of the baryons, but is transferred to internal energy. Otherwise, the temperature of the baryons is expected to be inaccurate.

We note that even if the internal energy of the baryons is significantly off this must not imply that the heat exchange between the DM and baryonic component is off as well. This is because the artificial small scale turbulence, i.e.\ the velocities of the numerical particles representing the baryons is present in calculations for the DM-baryon interactions. However, it is no longer possible to clearly distinguish between turbulence and temperature of the baryons on small scales.

For the test cases in Sect.~\ref{sec:test_problems}, we explicitly checked the kinetic energy of the baryonic particles to understand how strong the numerically induced turbulence is. In general, the strength of the required viscosity to dampen the turbulence depends on the strength of the DM-baryon interactions.
If artificial viscosity is used to dampen the numerically induced turbulence, it might need to only act on the smallest resolved flows, as larger scales are not likely to be affected by this numerical artefact.

Another relevant aspect is that the interactions artificially increase the temperature variation among the numerical baryonic particles. Thus, it could be favourable to have heat conduction that reduces this variation and prevents the internal energy of particles from approaching zero. Furthermore, it helps prevent negative internal energies (see also Sect.~\ref{sec:negative_energy_problem}) and thus makes the numerical scheme more robust.

The problems described above limit the physical cases to which our numerical scheme can be applied. Astrophysical gases may not in general have a physical viscosity and heat conduction being large enough to mitigate the described issues when needed. Artificial viscosity and artificial heat conduction can be applied instead if they allow for approximating the underlying physical system well enough, for example, ensuring they do not destroy relevant physical turbulence.

\subsection{Implementation}

We implemented the scheme for IDM in the cosmological $N$-body code \textsc{OpenGadget3}, a successor of \textsc{Gadget-2} described by \cite{Springel_2005}. The SIDM implementation in \textsc{OpenGadget3} was written by \cite{Fischer_2021a}, and is the basis for our implementation of IDM.
Further improvements to the SIDM module are described by \cite{Fischer_2021b, Fischer_2022, Fischer_2024a}.
We generalised the module and re-used it for the DM-baryon scattering.
In consequence, we employ the same scattering routine which has the advantage of being capable of simulating the limit of very anisotropic cross-sections \citep[fSIDM, see][]{Kahlhoefer_2014, Fischer_2021a}.
However, also we took advantage of the explicit energy conservation in parallel computations for the scattering of the particles.
The details of the parallelisation scheme, realised with the message passing interface (MPI), are explained in\ \cite{Fischer_2024a}.

The domain decomposition and the neighbour search in \textsc{OpenGadget3} have been described by \cite{Ragagnin_2016}.
Furthermore, the simulation code offers an SPH implementation \citep{Beck_2016} and an MFM scheme \citep{Groth_2023}.
We have implemented the interaction for both fluid schemes.
This allowed us to study the differences between SPH and MFM and evaluate which is better suited for our purpose.
In practice, the code does not save the internal energy but the entropy, this implies that we have to convert between entropy and internal energy as required.

For the scattering, we used a kernel assigned to each particle.
The kernel size of the DM particles is determined by searching for the next $N_\mathrm{ngb, DM}$ neighbours.
We search only for DM particles and ignore all other particle types.
In the case of the baryonic particles, the scheme to model hydrodynamics already employs a kernel size, $h_\mathrm{hydro}$, based on a neighbour number, $N_\mathrm{ngb, hydro}$. However, $h_\mathrm{hydro}$ might be larger or smaller than what we preferably use for the DM-baryon interactions.
Instead of performing another neighbour search, we rescale the kernel sizes for the interactions as we describe in the subsequent section (Sect.~\ref{sec:rescaling_kernel_size}).
These rescaled values are employed in Eq.~\eqref{eq:overlap} to compute the kernel overlap integral.
In our simulations, we used $N_\mathrm{ngb, DM} = 64$ for the DM particles, $N_\mathrm{ngb, SPH} = 230$ for the SPH particles and $N_\mathrm{ngb, MFM} = 32$ for the MFM particles.
The first number was motivated by an aim to obtain a sufficiently accurate estimate of the local DM density, so that the kernel size rescaling will work as expected. It is likely that a somewhat smaller number is sufficient as well. The numbers for the hydrodynamic schemes are chosen in the range of commonly employed values, which are known to provide accurate results.

The adaptive time-stepping scheme in \textsc{OpenGadget3} leads to active and passive particles, with the status of a particle being dependent on the time step that is computed. A detailed description can be found in the \textsc{Gadget-2} paper \citep{Springel_2005}.
Given that not only active particles interact with each other, but also active-passive particle pairs are possible, the search for scattering partners must be performed from both directions.
This means that we take all active DM particles and search for baryonic scattering partners, including active and passive particles.
As we might have missed pairs consisting of an active baryonic particle and a passive DM particle, we perform another search around all active baryonic pairs to find DM particles.
We note that the simulations presented in this paper employ a fixed time step for all particles implying that all particles are always active. The bidirectional search described above implies that we find every pair consisting of two active particles twice. However, for the interactions, we consider them only once per time step. This is realised with a criterion based on a unique identification number assigned to every particle.

\subsubsection{Rescaling of the local kernel size} \label{sec:rescaling_kernel_size}

In our IDM simulations, we have two species that are interacting with each other, this makes controlling the number of interactions between the numerical particles more complicated than in the single species case usually being present in SIDM studies.
The number of interactions does not simply depend on the kernel sizes chosen for baryons ($h_\mathrm{bary}$) and DM ($h_\mathrm{DM}$), but also on the local number density of the numerical particles of each species, namely,\ $n_\mathrm{bary}$ and $n_\mathrm{DM}$. The expected maximum number of interactions that a particle could undergo per time step can be expressed as,
\begin{equation}
    N = \frac{4}{3} \uppi \, (h_\mathrm{bary} + h_\mathrm{DM})^3 \, \max(n_\mathrm{bary}, n_\mathrm{DM}) \,.
\end{equation}
To effectively control the performance of the simulations it is important to control $N$. To do so, we introduce a factor, $\xi$, to obtain a rescaled kernel size of $h^* = h \, \xi$. We note that every numerical particle has its individual factor $\xi$. Altering the kernel sizes via $\xi$ allows us to control $N$, our target value is $N_\mathrm{idm}$.
Initially we assume $\xi = 1$ and update $\xi$ every time step,
\begin{equation} \label{eq:kernel_size_factor}
\xi' = \frac{1}{2 h} \sqrt[3]{\frac{N_\mathrm{idm}}{n_\mathrm{max}}} \,,\quad n_\mathrm{max} = \max\left(\frac{N_\mathrm{ngb, own}}{h^3}, \frac{N_\mathrm{ngb, other}}{{h_\mathrm{min}}^3}\right) \,.
\end{equation}
Here, we employ the minimum kernel size, $h_\mathrm{min}$, that a particle has seen over the last time step.
Moreover, $N_\mathrm{ngb, own}$, refers to the number of neighbours that were used to set the kernel size, $h$, of the species under consideration, i.e.\ the one to which the particle for which we want to compute $\xi'$ belongs. Similarly, $N_\mathrm{ngb, other}$, is the number of neighbours that were used for the other species and is relevant for $h_\mathrm{min}$.
We note that we derived Eq.~\eqref{eq:kernel_size_factor} by assuming that $h^*$ does only depend on the location but not on the species, namely,\ particles of different species should have the same rescaled kernel size if they are at the same location.

In practice, we employ the kernel size rescaling based on the DM and SPH/MFM kernel sizes.
It allows us to significantly speed up the simulations, depending on the chosen value for $N_\mathrm{idm}$.
Importantly, the rescaled kernel sizes are also employed in the search for scattering partners.

A reasonable choice for $N_\mathrm{idm}$ might be somewhat larger than typical neighbour numbers used in SIDM simulations. We note, from Eq.~\eqref{eq:kernel_size_factor} it follows for the single species case that $N = 8 \, N_\mathrm{ngb}$. This may give an idea of how to compare $N_\mathrm{idm}$ to the neighbour numbers employed in SIDM simulations. For our simulations we use $N_\mathrm{idm}=384$ if not stated otherwise.
This number is motivated by typical choices for SIDM. In addition, we tested to check that a larger number does not lead to a significant improvement, as described in Appendix~\ref{sec:convergence}.

\subsubsection{Time step criterion} \label{sec:time_step}
Analogously to time step criteria for SIDM \citep{Fischer_2021b, Fischer_2024a}, we derive the IDM time step criterion based on Eq.~\eqref{eq:probability} and Eq.~\eqref{eq:drag_force} with the aim to limit the interaction probability or respectively the fractional velocity change.
In the case of a velocity-independent cross-section, the interaction probability and the fractional velocity change increase monotonically with relative velocity.
Consequently, we estimate the maximal scattering velocity that each particle could encounter.
Therefore we compute for every particle, $v_{\mathrm{max}} = \max(v + v_\mathrm{cut})$, by taking the maximum over all pairs. 
In the case of a velocity-dependent cross-section, there might eventually exist a finite velocity $v_\mathrm{e}$ for which the interaction probability and fractional velocity change become maximal.
This is for example the case for the velocity-dependent cross-section employed by \cite{Fischer_2024a}.\footnote{The same velocity-dependence has previously been used by many other authors in the context of SIDM.}
In this case we can combine $v_\mathrm{max}$ and $v_\mathrm{e}$ to obtain a critical velocity $v_\mathrm{c} = \min(v_\mathrm{max}, v_\mathrm{e})$ to formulate the time step criteria. For a velocity-independent cross-section, we can simply set $v_\mathrm{c} = v_\mathrm{max}$.

The time step criterion for the frequent interactions, that is small-angle scattering, depends on the momentum transfer cross-section (see Eq.~\eqref{eq:transfer_cross-section}) and is given by
\begin{equation}
    \Delta t_{\mathrm{freq}} = \frac{\tau_\mathrm{freq}}{m_\mathrm{DM} \, v_c \, \Lambda_{ix}} \, \frac{\mu}{f_\mathrm{bary}} \, \left(\frac{\sigma_\mathrm{T}(v_c)}{m_\chi}\right)^{-1} \,.
\end{equation}
Analogously, we can give the time step criterion for rare interactions, this is large-angle scattering, employing the total cross-section,
\begin{equation}
    \Delta t_{\mathrm{rare}} = \frac{\tau_\mathrm{rare}}{m_\mathrm{DM} \, v_c \, \Lambda_{ix}} \, \frac{\mu}{f_\mathrm{bary}} \, \left(\frac{\sigma(v_c)}{m_\chi}\right)^{-1} \,.
\end{equation}
The factors $\tau_\mathrm{freq}$ and $\tau_\mathrm{rare}$, allow the size of the time step to be controlled, namely, by limiting the fractional velocity change and the interaction probability, respectively.
In contrast to the SIDM time step criterion employed by \cite{Fischer_2024a}, we do not use the self-overlap $\Lambda_{ii}$. In the case of two separate species when the kernel size is determined using particles of the same species only, it can happen that the kernel sizes differ between the species vastly, leading to a dramatic over or underestimate of the time step when using $\Lambda_{ii}$. Instead, we employ $\Lambda_{ix}$ when computing the time step of particle $i$, where $x$ refers to the particle with the smallest kernel size that particle $i$ has seen over the last time step. For the computation of $\Lambda_{ix}$, however, we assume that both particles are located at the same position; this is the same as for the self-overlap but with two different kernel sizes\footnote{When using the kernel size rescaling described in Sect.~\ref{sec:rescaling_kernel_size}, the use of $\Lambda_{ii}$ should be unproblematic. However, we nevertheless use $\Lambda_{ix}$.}.

\subsubsection{Considerations of the energy conservation} \label{sec:energy_conservation}
Although the DM-baryon interactions are formulated in a strict energy-conserving manner, this does not necessarily imply that they do not harm energy conservation. While SPH and MFM can be formulated in a strict conserving manner as well, the combination of an explicit conservative hydrodynamics scheme with IDM might result in a loss of conservation of total energy.
The IDM scheme changes the velocity and in particular the internal energy of the baryonic particles. Within pure hydro schemes, those quantities are not expected to be altered at the stage where the IDM kicks happen. This can invalidate previously computed hydrodynamical accelerations and cause energy non-conservation.

The IDM case is much more complicated than the one of SIDM. In the latter, we only have to deal with the DM and its accelerations are computed from gravity only. In consequence, the DM acceleration does not contain any velocity-dependent term\footnote{Although the integration scheme we use does not require velocity-dependent terms for gravity, we want to note that this is not true for all schemes. For example, the fourth-order Hermite integration method is based on higher-order derivatives introducing a velocity-dependent term \citep{Makino_1992}.}, which would be affected by the SIDM kicks, but is purely based on the positions of the DM particles. In contrast, for IDM we might have velocity-dependent terms such as viscosity in the hydrodynamics scheme; importantly, however, the hydrodynamical accelerations depend on the internal energy, which is modified by the DM-baryon interactions.

As a consequence, it matters where in the time integration scheme the IDM interactions are executed. In \textsc{OpenGadget3}, we use a leapfrog scheme in the kick-drift-kick (KDK) formulation. For the simulations in Sect.~\ref{sec:test_problems} and Sect.~\ref{sec:halo_formation}, we implemented the IDM interactions between the two half-step kicks. But given that we alter the internal energy of the baryons we invalidate the acceleration used in the second half step-kick. As a consequence, energy is not explicitly conserved any more. We note, that the issue of energy conservation might be more complicated than what we describe here; for example additional complications may arise from using variable time steps and a wake-up scheme employed as part of the solver for hydrodynamics \citep[e.g.][]{Saitoh_2009, Pakmor_2012}. We leave a detailed investigation for future work. 
Nevertheless, it is in principle possible to improve on energy conservation by modifying the time integration while accepting higher computational costs as we discuss and show in Appendix~\ref{sec:improved_energy_conservation}. Though for the test problems we simulated (see Sect.~\ref{sec:test_problems}) this has not been of much relevance as the energy error is small enough to hardly affect the agreement between the numerical and exact solution.

\section{Test problems} \label{sec:test_problems}
In this section, we study multiple test problems, including heat conduction and momentum transfer between baryons and DM.
Firstly we consider a set-up where heat flows from the DM to the baryons and one with the opposite case where heat flows from baryons to DM (Sect.~\ref{sec:test_problems_heat_conduct}).
Secondly, we study a problem in which DM and baryons initially move relative to each other and exchange momentum (Sect.~\ref{sec:test_problems_momentum_transfer}).
While we first limit those tests to equal mass ratios and small-angle scattering, we also test the implementation for an unequal mass ratio (Sect.~\ref{sec:test_problems_unequal_mass}) and for large scattering angles, namely isotropic scattering (Sect.~\ref{sec:test_problems_isotropic}).

In this work, we consider two differential cross-sections, a forward-dominated model and an isotropic one. The two cross-sections are velocity-independent and the interactions are elastic. The forward-dominated cross-section is given by the limit where the transfer cross-section, $\sigma_\mathrm{T}$ (Eq.~\eqref{eq:transfer_cross-section}), is held constant while the scattering angles approach zero. This model could also be expressed as
\begin{equation}    \label{eq:cross-section_fwd}
    \left.\frac{\mathrm{d}\sigma}{\mathrm{d}\Omega_\mathrm{cms}}\right|_\mathrm{fwd}=\lim_{\epsilon\to 0} \frac{\sigma_\mathrm{T}}{8 \, \uppi\ln(\epsilon^{-2})} \frac{1}{\left(\epsilon^2 + \sin^2{\theta_\mathrm{cms}/2}\right)^2} \,.
\end{equation}
A similar formulation but for identical particles has been used in several studies of SIDM, usually referred to as frequent scattering \citep[e.g.][]{Kahlhoefer_2014, Kummer_2019, Sabarish_2024, Arido_2025}.
The isotopic model can be formulated with the total cross-section, $\sigma$,
\begin{equation}    \label{eq:cross-section_iso}
    \left.\frac{\mathrm{d}\sigma}{\mathrm{d}\Omega_\mathrm{cms}}\right|_\mathrm{iso}= \frac{\sigma}{4 \uppi} \, .
\end{equation}
and would for example follow from hard-sphere scattering.

\subsection{Heat conduction} \label{sec:test_problems_heat_conduct}

We simulate DM and baryons together, while both species interact with each other. However, at the same time we neglect gravitational forces.
The test set-up for our heat conduction problem consists of a cube with a side length of 10 kpc and periodic boundary conditions.
Both components initially have the same density and are at rest.
The velocities of the DM component follow a Maxwell-Boltzmann distribution. The baryonic particles have no bulk velocity but a non-zero temperature.
The one-dimensional velocity dispersion of the DM particles is $\nu = 2 \, \mathrm{km} \, \mathrm{s}^{-1}$ and the baryons have an internal energy that corresponds to $10\%$ of the kinetic energy of the DM.
We have $10^5$ DM particles and $46656$ baryonic particles arranged in a lattice with $36$ particles per dimension.
The particle numbers are chosen such that the baryonic particles are roughly twice as massive as the DM particles.
Each component accounts for a mass of $10^{10} \, \mathrm{M_\odot}$.
This implies a DM density, $\rho_\mathrm{DM} = 10^7 \, \mathrm{M_\odot} \, \mathrm{kpc}^{-3}$ and a baryonic density, $\rho_\mathrm{bary} = 10^7 \, \mathrm{M_\odot} \, \mathrm{kpc}^{-3}$, while the energy densities are $w_\mathrm{DM} = 6 \times 10^{7} \,\mathrm{M_\odot} \, \mathrm{km}^2 \, \mathrm{s}^{-2} \, \mathrm{kpc}^{-3}$ and $w_\mathrm{bary} = 6 \times 10^6 \, \mathrm{M_\odot} \,\mathrm{km}^2 \, \mathrm{s}^{-2} \, \mathrm{kpc}^{-3}$. To compute the DM-baryon interaction we employ a cubic spline kernel \citep{Monaghan_1985}. It is used in Eq.~\eqref{eq:overlap} to compute the kernel overlap relevant for the interaction probability (Eq.~\eqref{eq:probability}) and strength of the drag force (Eq.~\eqref{eq:drag_force}).
We simulate the set-up with an extremely anisotropic cross-section (Eq.~\eqref{eq:cross-section_fwd}) and use a velocity-independent momentum transfer cross-section of $\sigma_\mathrm{T} / m_\chi = 10\,\mathrm{cm}^2 \, \mathrm{g}^{-1}$.
In addition, we use the MPI parallelisation scheme for all tests of this Section.

\begin{figure}
    \centering
    \includegraphics[width=\columnwidth]{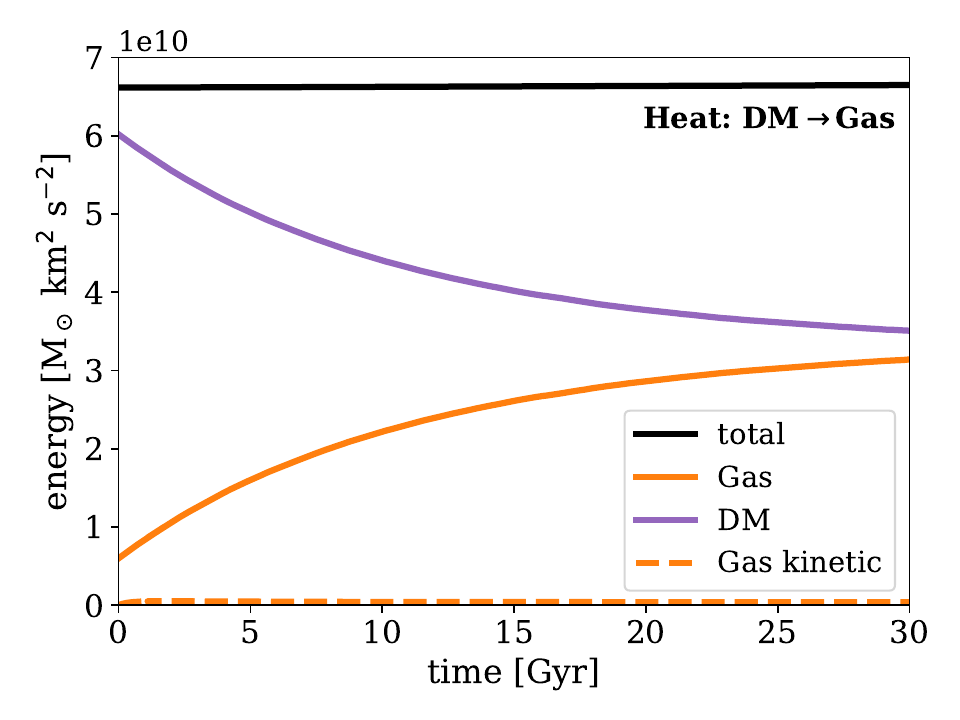}
    \caption{Heat conduction problem where energy flows from the dark matter to the baryons. Different types and components of the energy are shown as a function of time. The total energy of the system is illustrated in black. For DM the energy is shown in violet, it corresponds to the kinetic energy of the particles as other contributions are zero. In orange, we illustrate the energy of the baryons. It consists of the internal or thermal energy and the kinetic energy (bulk motion). The kinetic part is also displayed separately (dashed line). The simulation was conducted employing a forward dominated cross-section and SPH to describe the gas (baryons).}
    \label{fig:results_heat_dm2gas}
\end{figure}

In Fig.~\ref{fig:results_heat_dm2gas}, we show the results for the heat conduction problem when using SPH with 230 neighbours and a Wendland $C^6$ kernel \citep{Dehnen_2012}.
Neighbour numbers of this size are a typical choice for our SPH implementation and have proven to give reliable results.
The DM neighbour number is $N_\mathrm{ngb, DM}=64$ and the interaction number is $N_\mathrm{idm}=384$.
The first one is chosen large enough to give a sufficiently accurate estimate for the local DM density and is used for the kernel size rescaling only. The second one controls the number of particles that a given particle could interact with per time step. In Appendix~\ref{sec:convergence}, we show that increasing the number any further does not significantly improve the results.
Moreover, we employed artificial viscosity and artificial head conduction \citep{Price_2012, Beck_2016}.
The artificial viscosity is formulated to act only against high-velocity divergence, aiming to leave rotating or shearing flows unchanged. The settings were chosen to be the same as those for the \textsc{Magneticum} simulations \citep{Dolag_2025}. Hence, we do not expect that they would harm the formation of galaxies, for example, affecting the star formation rate.
Furthermore, we note that the same fixed time step of $\Delta t_\mathrm{SPH} = 0.024 \,\mathrm{Gyr}$ is employed for all particles, this is true for all SPH simulations of this section.
For the size of the time step, the value implied by the time step criterion for the hydrodynamics is taken as an orientation \citep{Beck_2016}. A smaller choice for the time step could help to improve the results, as we show in Appendix~\ref{sec:convergence}.
From the figure, we can see that the total energy (black) stays constant, but the energy of the DM particles (violet) is decreasing over time and the energy of the baryons (orange) is increasing.
We note, that we separately show the kinetic energy from the bulk motion of the baryons, i.e.\ the kinetic energy of the numerical particles.
It is supposed to be zero and as we can see, it increases only slightly over the course of the simulation.

The set-up for the heat flow from baryons to DM is the same as for the heat conduction from the DM to the baryons, but here we interchange the energies, i.e.\ the baryons contain 10 times the energy of the DM. We simulate the set-up with the same cross-section as before ($\sigma_\mathrm{T} / m_\chi = 10\,\mathrm{cm}^2 \, \mathrm{g}^{-1}$) and thus expect the thermalisation to take place at the same speed. 

\begin{figure}
    \centering
    \includegraphics[width=\columnwidth]{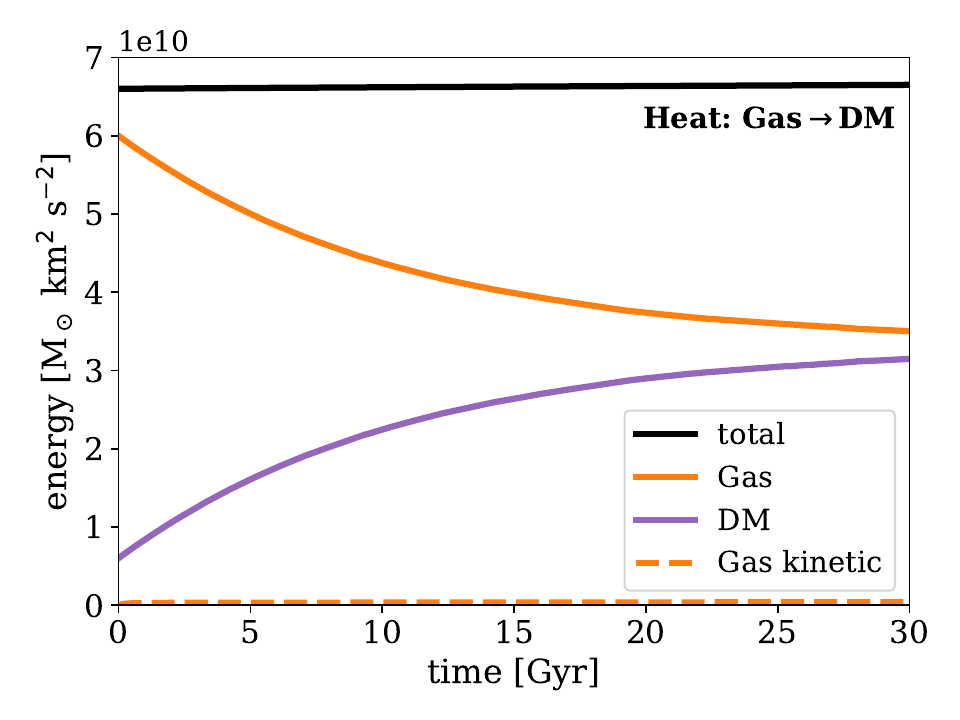}
    \caption{Same as in Fig.~\ref{fig:results_heat_dm2gas}, but for the heat conduction problem with energy flowing from the baryons to the DM.}
    \label{fig:results_heat_gas2dm}
\end{figure}

\begin{figure}
    \centering
    \includegraphics[width=\columnwidth]{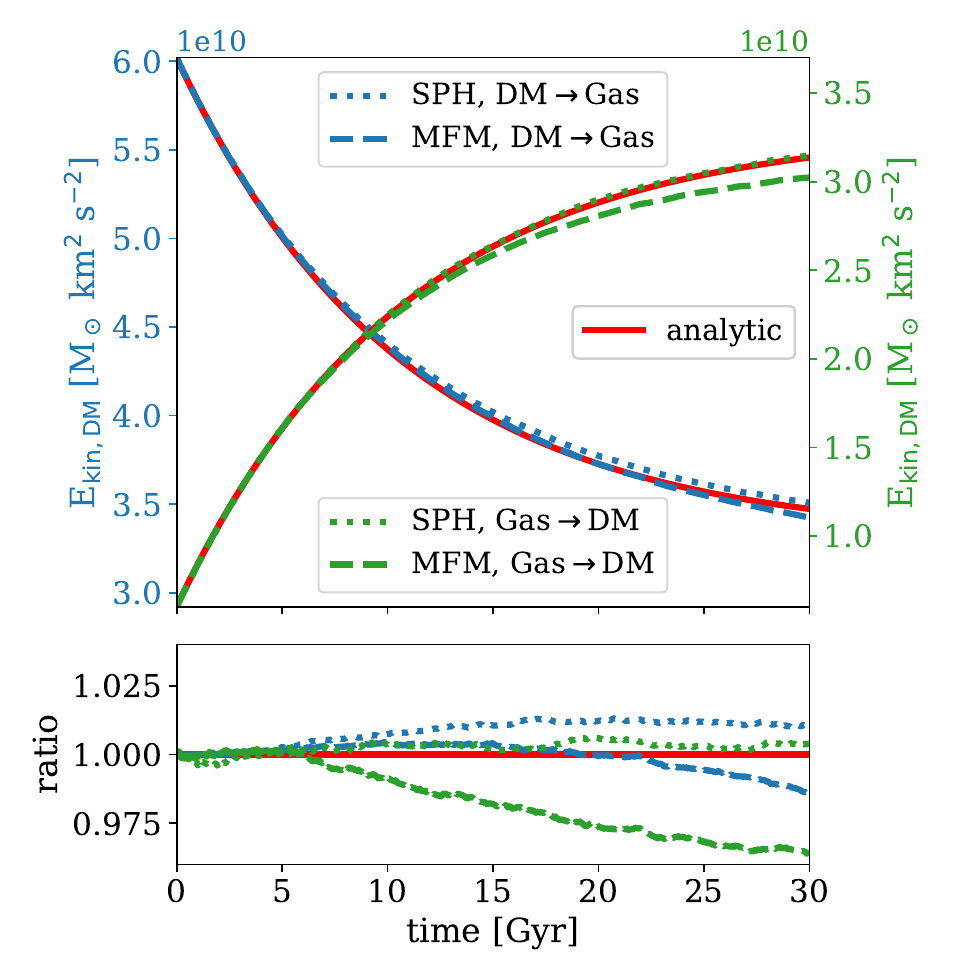}
    \caption{Kinetic energy of the DM as a function of time for both heat conduction set-ups. For heat flowing from the DM to baryons (DM$\rightarrow$Gas) the results are given in blue and for heat flowing from baryons to DM (Gas$\rightarrow$DM) in green. We display the results for both methods SPH (dotted) and MFM (dashed). The red lines give the exact solution according to Eqs.~\eqref{eq:heat_conduct_exact_solution} and~\eqref{eq:heat_conduct_kappa}. The upper panel gives the absolute values, and the lower panel displays the ratios to the exact solution.}
    \label{fig:results_heat_dmkineticenergy}
\end{figure}

\begin{figure}
    \centering
    \includegraphics[width=\columnwidth]{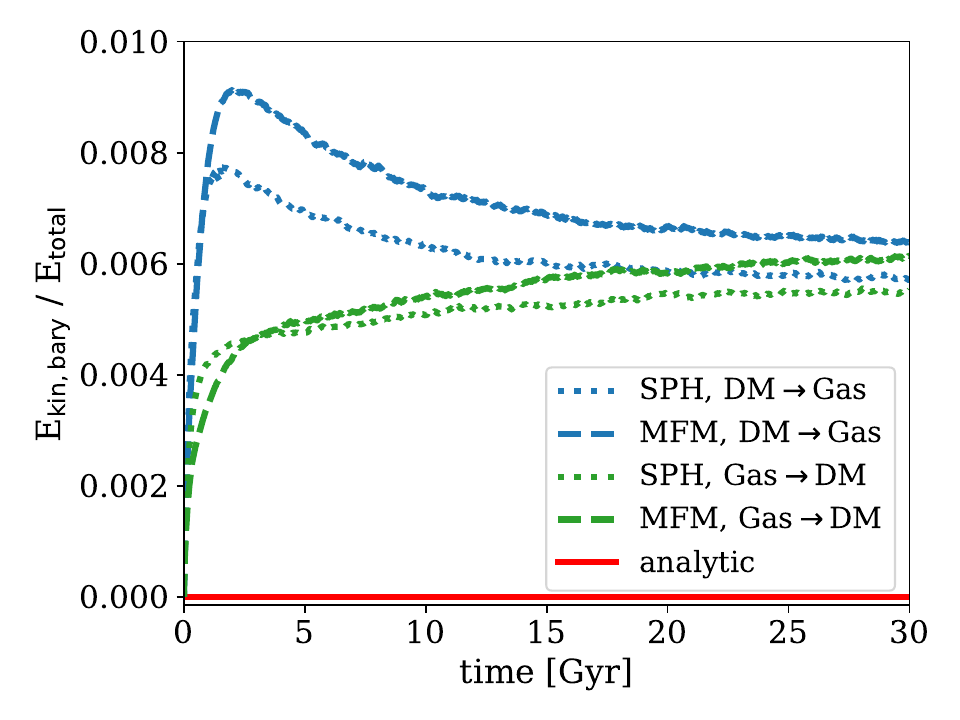}
    \caption{Kinetic energy of the baryonic particles in units of the total energy is shown as a function of time. Labels, colours, and line types are the same as in Fig.~\ref{fig:results_heat_dmkineticenergy}. We note that this figure is basically a zoom-in on the orange dashed curves displayed in Figs.~\ref{fig:results_heat_dm2gas} and~\ref{fig:results_heat_gas2dm}.}
    \label{fig:results_heat_barykineticenergy}
\end{figure}

In Fig.~\ref{fig:results_heat_gas2dm}, we show the results for the problem with heat conduction from the baryons to the DM using the SPH implementation. Here the energy densities of DM and baryons are swapped compared to the previous set-up.
We can see that the thermalisation takes place at roughly the same speed as for the set-up with heat conduction in the other direction.
Any difference in the thermalisation speed is not physical but due to numerical errors, a comparison to the exact solution follows next with Fig.~\ref{fig:results_heat_dmkineticenergy}.
Moreover, the kinetic energy of the baryonic particles stays small over the course of the simulation, a detailed investigation follows later with Fig.~\ref{fig:results_heat_barykineticenergy}.

In the following, we compare the test simulations in larger detail.
We did not only use SPH but also ran the same simulations using MFM with 32 neighbours. With this, we follow a common choice for MFM simulations \citep[e.g.][]{Gaburov_2011}. Moreover, we employ the cubic spline kernel \citep{Monaghan_1985} for MFM and use Eq.~\eqref{eq:kernel_size_factor} to rescale the kernel size for the DM-baryon interactions as we do for SPH too, while keeping the DM neighbour number at 64 as before. 
We note, that for MFM we do not employ artificial viscosity and artificial heat conduction, the MFM scheme itself already gives rise to numerical viscosity.
In addition, we employ a fixed time step of $\Delta t_\mathrm{MFM} = 0.006 \,\mathrm{Gyr}$ for all particles. This is a quarter of the value for the SPH simulations and used for all MFM simulations of this section. Again, we took the corresponding time step criterion as an orientation \citep{Groth_2023}. Decreasing the time step can help to improve the results as we show in Appendix~\ref{sec:convergence}.

In Fig.~\ref{fig:results_heat_dmkineticenergy}, we show the kinetic energy of the DM particles.
Depending on the set-up it is increasing or decreasing over time.
We find that a significant difference between SPH and MFM is present which is growing over time.
The energy of the DM particles in the MFM runs is always lower than in the corresponding SPH runs, no matter in which direction the heat is flowing.
This is related to energy non-conservation in the MFM runs as we see later.

Figure~\ref{fig:results_heat_dmkineticenergy} also allows us to compare the simulation results to the exact solution indicated by the red lines. The analytic description follows from \cite{Dvorkin_2014} \citep[see also][]{Munoz_2015} and simplifies in our case with a velocity-independent cross-section to
\begin{equation} \label{eq:heat_conduct_exact_solution}
    E_\mathrm{DM}(t) = E_\mathrm{DM, eq} + (E_\mathrm{DM, ini} - E_\mathrm{DM, eq}) \, e^{- \kappa t} \,.
\end{equation}
Here, $E_\mathrm{DM, ini}$ is the initial energy of the DM and $E_\mathrm{DM, eq} = (E_\mathrm{tot} \, r) / (1+r)$, with $E_\mathrm{tot}$ being the total energy (i.e.\ the sum of DM and baryons). The speed at which the heat transfer happens is set by
\begin{equation} \label{eq:heat_conduct_kappa}
    \kappa = \frac{8}{\sqrt{\uppi}} \, \left(\frac{2}{3}\right)^{3/2} \frac{\sqrt{w_\mathrm{tot} \, \rho_\mathrm{tot}}}{1+r} \, \frac{\sigma_\mathrm{T}}{m_\chi} \,.
\end{equation}
The total energy density $w_\mathrm{tot} = w_\mathrm{DM} + w_\mathrm{bary}$ is given by the sum of the energy density (i.e.\ energy per volume) of DM and baryons. Analogously the total matter density is $\rho_\mathrm{tot} = \rho_\mathrm{DM} + \rho_\mathrm{bary}$.
We note that the equations above are only valid when $\rho_\mathrm{DM} = \rho_\mathrm{bary}$, given that we made this assumption for obtaining a simpler expression (a more general formulation can be found in Appendix~\ref{sec:como_test}).

Overall, the simulation results agree well with the exact solution, in particular the SPH runs (see Fig.~\ref{fig:results_heat_dmkineticenergy}).
The MFM scheme agrees very well at the beginning of the simulation and gives a bit too low energies at the late stages of the evolution.
The coupling of IDM to the hydro schemes can lead to different numerical artefacts affecting how well the simulation results agree with the exact solution. In the following, we look closer into this.

Next, we study the kinetic energy of the baryonic particles. It is supposed to stay zero as DM and baryons are at rest and the interactions only heat up or cool down the baryons.
However, in practice, our numerical scheme must allow for an exchange of momentum in each DM-baryon interaction.
In the convergence limit, the kinetic energy of the baryons would stay zero for our heat conduction test problems.
How much it deviates from zero can only (in a limited sense) be considered a measure of how accurate the simulations are (see Sect.~\ref{sec:viscosity_and_heat_conduction}).
In Fig.~\ref{fig:results_heat_barykineticenergy}, the kinetic energy of the baryonic particles is shown.
It is visible that initially, the energy is increasing steeply for all runs.
For the simulation with heat flow from the baryons to DM (green) a plateau is reached after the sharp increase and the SPH and MFM runs exhibit a very similar behaviour. In general, the difference between the simulated set-ups is larger then between the numerical schemes for modelling hydrodynamics.
The simulations for the set-up with heat flow from DM to baryons reaches a peak about twice as high as for the test with the heat flow in the opposite direction. After reaching the peak, the kinetic energy of the baryonic particles declines and reaches almost the value of the other simulations. Here, viscous forces reduce the artificial small-scale turbulence. 

\begin{figure}
    \centering
    \includegraphics[width=\columnwidth]{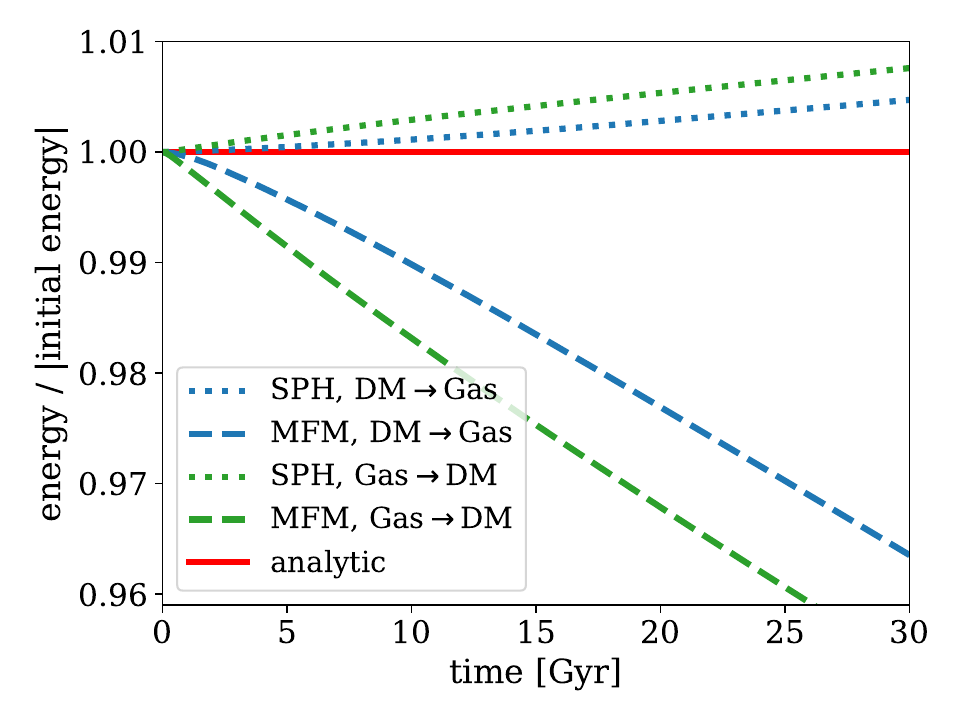}
    \caption{Total energy in units of the initial energy as a function of time to study energy conservation.
    The results are displayed for the two heat conduction set-ups and both fluid methods, SPH, and MFM.
    Labels, colours and line types are chosen as in Fig.~\ref{fig:results_heat_dmkineticenergy}.}
    \label{fig:results_heat_energy_conservation}
\end{figure}

The last aspect we studied with our simulations for the heat conduction problem is energy conservation. In Fig.~\ref{fig:results_heat_energy_conservation}, we show how well total energy is conserved as a function of time. Ideally, we expect energy to be perfectly conserved as indicated by the red line. However, in practice, it is subject to numerical error. We can see that the energy errors are monotonically increasing over the simulation time. Interestingly, the errors are much smaller for SPH compared to MFM, where for the later one they rise up to a few percent of the total energy. As a consequence, the SPH implementation appears to be preferable in terms of energy conservation, with energy errors below one percent. In Sect.~\ref{sec:energy_conservation}, we discuss reasons for non-conservation, and in Appendix~\ref{sec:improved_energy_conservation}, we demonstrate that the conservation of total energy can be improved for the case of SPH. Moreover, we ran tests varying the time step size and the interaction number $N_\mathrm{idm}$, they are presented in Appendix~\ref{sec:convergence}.

Lastly, we want to mention that we performed a test with a similar set-up in an expanding space. In Appendix~\ref{sec:como_test}, we show the results and demonstrate that we can also accurately model IDM when using comoving integration.

\subsection{Momentum transfer} \label{sec:test_problems_momentum_transfer}

We considered a problem similar to the previous one, but now the DM and baryonic components are moving relative to each other and contain the same energy.
Their matter densities are the same as in Sect.~\ref{sec:test_problems_heat_conduct} ($\rho_\mathrm{DM} = \rho_\mathrm{bary}= 10^7 \, \mathrm{M_\odot} \, \mathrm{kpc}^{-3}$).
The one-dimensional velocity dispersion of the DM is $\nu = 0.5 \, \mathrm{km} \, \mathrm{s}^{-1}$ and the relative velocity between the two components is $4 \, \mathrm{km} \, \mathrm{s}^{-1}$.
The corresponding energy densities (excluding the bulk motion) are $w_\mathrm{DM} = w_\mathrm{bary} = 3.75 \times 10^6 \, \mathrm{M_\odot} \, \mathrm{km}^2 \, \mathrm{s}^{-2} \, \mathrm{kpc}^{-3}$. Thus, the two components have the same energy in their common centre-of-mass frame. As a result of having an equal mass ratio ($r=1$), we do not expect any heat transfer to occur between the two components, but they should decelerate relative to each other and in consequence heat up, while the total energy of each component stays constant.
Again we consider velocity-independent forward dominated scattering with a momentum transfer cross-section per DM particle mass of $\sigma_\mathrm{T} / m_\chi = 10\,\mathrm{cm}^2 \, \mathrm{g}^{-1}$.

Based on the work by \cite{Dvorkin_2014}, we can analytically estimate the solution for this test problem assuming a velocity-independent cross-section. In the limit where the relative velocity is much larger than the velocity dispersion of the two components, the evolution can be well described by a drag force decelerating the DM and baryons. The drag deceleration acting on the DM component is given by
\begin{equation}
    a_\mathrm{drag} = - \frac{\rho_\mathrm{bary} \, v^2_\mathrm{rel}}{1 + r} \frac{\sigma_\mathrm{T}}{m_\chi} \, .
\end{equation}
In the regime where the velocity dispersion dominates over the relative velocity, the deceleration can be expressed as
\begin{equation}
    a_\mathrm{disp} = - \frac{8 \sqrt{2}}{3 \sqrt{\uppi}} \, \frac{\rho_\mathrm{bary} \, v_\mathrm{rel}}{1 + r} \, \sqrt{\nu^2_\mathrm{DM} + \nu^2_\mathrm{bary}} \, \frac{\sigma_\mathrm{T}}{m_\chi} \, G_0\left( \frac{v^2_\mathrm{rel}}{\nu^2_\chi + \nu^2_\mathrm{bary}} \right)\, .
\end{equation}
Here, $\nu^2_\mathrm{DM}$ and $\nu^2_\mathrm{bary}$ are the one-dimensional velocity dispersion of the DM and baryons respectively. The function $G_0$ can be expressed as a series,
\begin{equation}
    G_0(X) = 1 + \frac{X}{10} - \frac{X^2}{280} + \frac{X^3}{5040} + ...\, ,
\end{equation}
where the variable $X = v^2_\mathrm{rel} / (\nu^2_\chi + \nu^2_\mathrm{bary})$.
To estimate the deceleration we compute $G_0$ up to the third order and interpolate between the two regimes,
\begin{equation} \label{eq:momentum_transfer_analytic_estimate}
    \frac{\mathrm{d}v_\mathrm{DM}}{\mathrm{d}t} = a_\mathrm{drag} \, f(X) + a_\mathrm{disp} \, \left(1 - f(X)\right)
\end{equation}
For the interpolation, we used a logistic weighting function,
\begin{equation}
    f(X) = \frac{1}{1 - e^{a (1 - b \, X)}} \,.
\end{equation}
For the comparison, we employ the following parameters for the weight, $a = 3.0$ and $b = 0.1$.
Moreover, we note that in our set-up $\rho_\mathrm{DM} = \rho_\mathrm{bary}$ and thus the two components experience the same deceleration. To obtain the time evolution of the relative velocity between the two components we integrate Eq.~\eqref{eq:momentum_transfer_analytic_estimate} numerically.

For the simulations, we use the same numerical parameters as for the heat conduction problem and run the test with SPH and MFM. In particular, we employ the same
resolution ($N_\mathrm{DM} = 10^5$, $N_\mathrm{bary} = 46656$), the same neighbour numbers ($N_\mathrm{ngb, DM} = 64$, $N_\mathrm{ngb, SPH} = 230$, $N_\mathrm{ngb, MFM} = 32$, $N_\mathrm{idm} = 384$), the same kernel functions and the same fixed time step ($\Delta t_\mathrm{SPH} = 0.024 \, \mathrm{Gyr}$, $\Delta t_\mathrm{MFM} = 0.006 \, \mathrm{Gyr}$). Again, the SPH run is conducted using artificial viscosity and artificial heat conduction.

\begin{figure}
    \centering
    \includegraphics[width=\columnwidth]{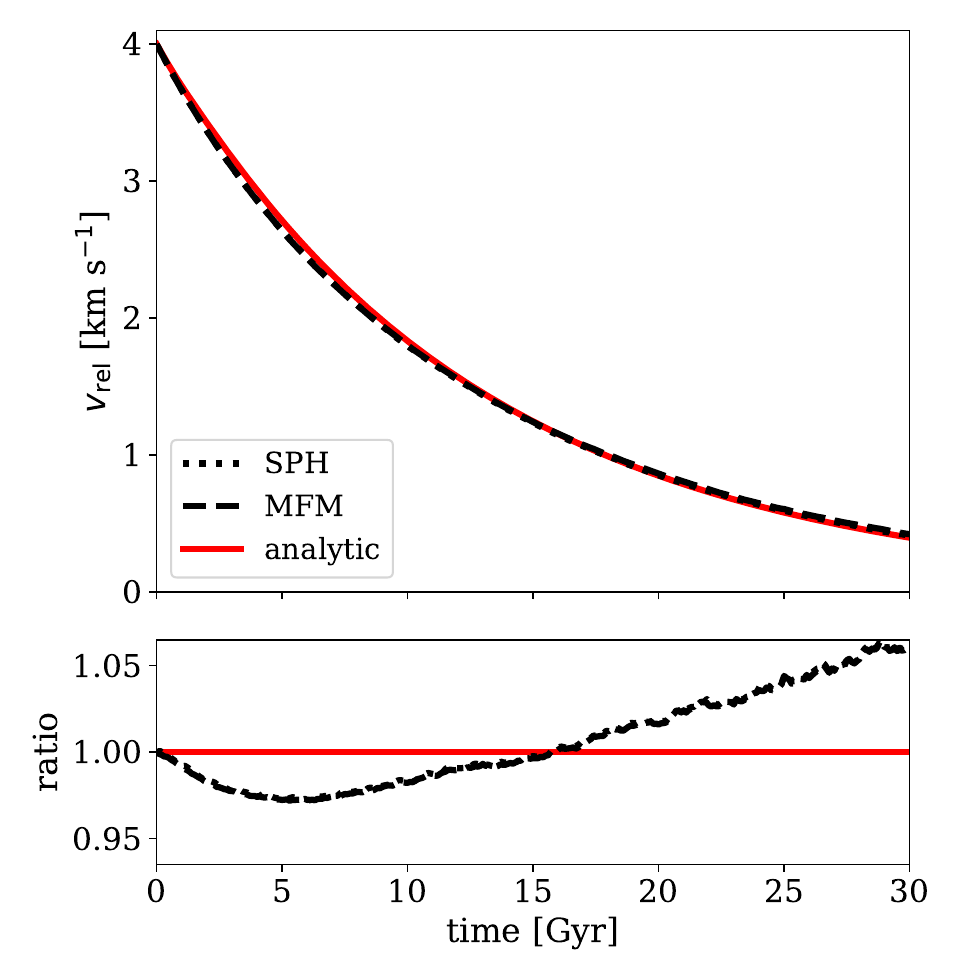}
    \caption{Relative velocity between DM and baryons for the momentum transfer problem. The time evolution for our SPH and MFM simulations (black) as well as the analytic estimate (red) of the relative velocity are displayed. The latter is computed according to Eq.~\eqref{eq:momentum_transfer_analytic_estimate}.
    The upper panel gives the absolute values, and the lower panel displays the ratios to the exact solution.
    We note, that the SPH and MFM results are very similar.}
    \label{fig:results_momentum_relvel}
\end{figure}

In Fig.~\ref{fig:results_momentum_relvel}, we display the relative velocity between the DM and the baryons as a function of time and compare it to our analytic estimate.
As expected the relative velocity decreases as the baryon-DM interactions lead to a momentum transfer between the two components. The simulation results agree well with the analytic estimate and the two hydro schemes SPH and MFM behave very similarly. We note that the accuracy of the analytic estimate, given that it is an interpolation of two regimes, might account for some of the deviation between numerical and analytic results.

\begin{figure}
    \centering
    \includegraphics[width=\columnwidth]{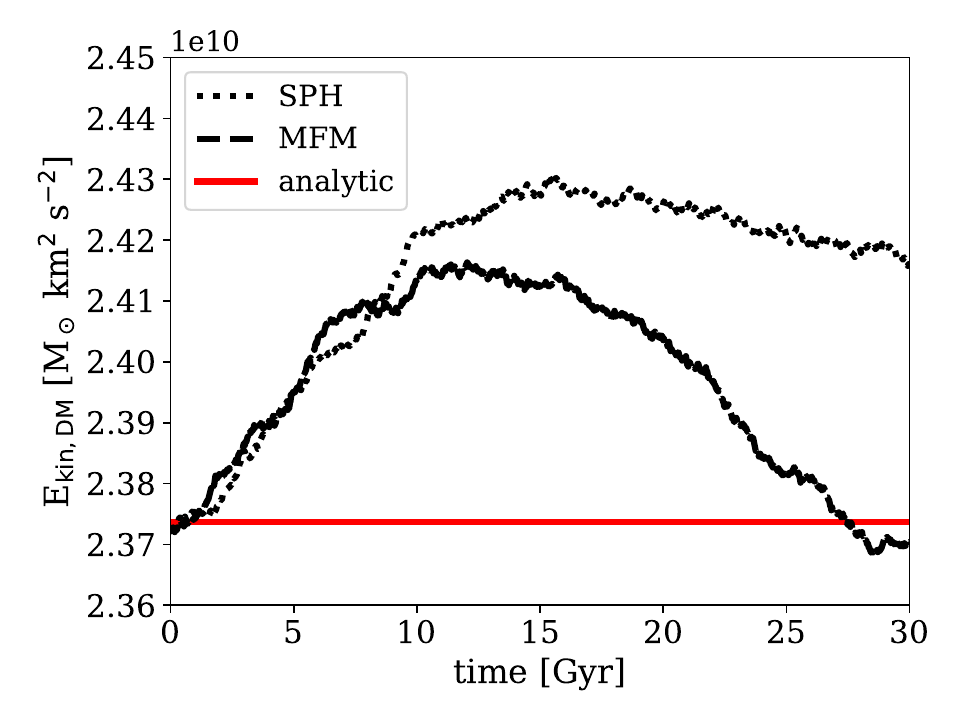}
    \caption{Kinetic energy of the DM particle as a function of time for the momentum transfer problem. The dotted lines give the results using SPH and the dashed ones are for MFM.}
    \label{fig:results_momentum_dmkineticenergy}
\end{figure}

We also show the kinetic energy of DM in Fig.~\ref{fig:results_momentum_dmkineticenergy}. It is expected to remain constant over time as the energy of the bulk motion is converted into random motion.
However, in the beginning, we see an increase in the kinetic energy, which is more pronounced for SPH.
Later, the DM significantly loses energy in the MFM run, which is related to energy non-conservation as we see next. In contrast, the kinetic energy of the DM for the SPH run stays rather constant or slightly decreases at this stage.

\begin{figure}
    \centering
    \includegraphics[width=\columnwidth]{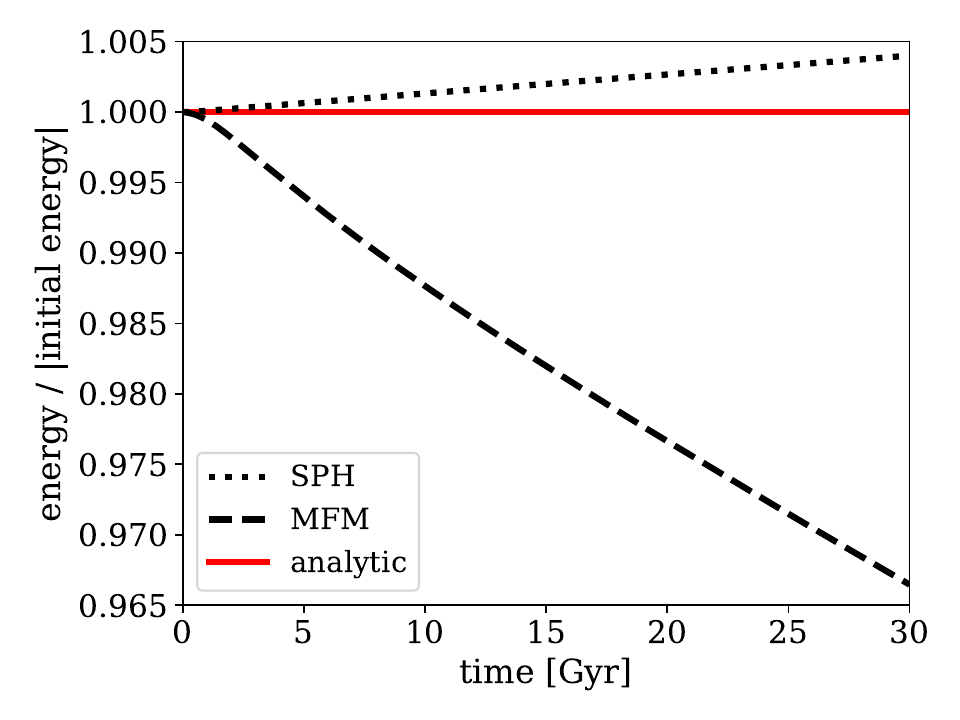}
    \caption{Energy conservation for the momentum transfer problem as a function of time. The total energy divided by the initial energy is given for SPH (dotted line) and MFM (dashed line).}
    \label{fig:results_momentum_energy_conservation}
\end{figure}

Finally, we studied the energy conservation for this test problem. In Fig.~\ref{fig:results_momentum_energy_conservation}, we can see that the energy is better conserved for SPH than for MFM. This is in line with our previous findings for the heat conduction problem illustrated in Fig.~\ref{fig:results_heat_energy_conservation}.

\subsection{Unequal mass ratio} \label{sec:test_problems_unequal_mass}

In addition to the tests where particles with an equal mass ratio scatter, we also consider a problem where the physical DM particle has only $1/1000$th of the mass of its baryonic scattering partner, i.e.\ $r=1000$.
Further we choose, $f_\mathrm{bary} = 1$ and each mass component makes up $10^{10} \, \mathrm{M_\odot}$.
We have an unequal number of baryonic and DM particles, $N_\mathrm{bary} = 46656$ and $N_\mathrm{DM} = 10^5$.
As for the other test problems, we have a periodic box with a side length of 10 kpc containing a constant density ($\rho_\mathrm{DM} = \rho_\mathrm{bary}= 10^7 \, \mathrm{M_\odot} \, \mathrm{kpc}^{-3}$).
The system is evolved at a fixed time step of $\Delta t = 0.015 \, \mathrm{Gyr}$.
Small-angle scattering is simulated and a momentum transfer cross-section per DM particle mass of $\sigma_\mathrm{T} / m_\chi = 1000 \, \mathrm{cm}^2 \, \mathrm{g}^{-1}$ is employed.
Initially, the two components contain the same energy, but due to the scattering with an unequal mass ratio energy is transferred from the baryons to the DM. In the equilibrium state, which is asymptomatically approached by the system, the DM has an energy that is $r$ times, namely,\ 1000 times, the energy of the baryons.
The initial energy of the DM is given by a one-dimensional velocity dispersion of $\nu = 2 \, \mathrm{km} \, \mathrm{s}^{-1}$.
This implies that the initial energy densities are $w_\mathrm{DM} = w_\mathrm{bary} = 6 \times 10^{7} \, \mathrm{M_\odot} \, \mathrm{km}^2 \, \mathrm{s}^{-2} \, \mathrm{kpc}^{-3}$; whereas in the equilibrium state, they become $w_\mathrm{DM} = 11.988 \times 10^{7} \, \mathrm{M_\odot} \, \mathrm{km}^2 \, \mathrm{s}^{-2} \, \mathrm{kpc}^{-3}$ and $w_\mathrm{bary} = 0.012 \times 10^{7} \,\mathrm{M_\odot} \, \mathrm{km}^2 \, \mathrm{s}^{-2} \, \mathrm{kpc}^{-3}$.

\begin{figure}
    \centering
    \includegraphics[width=\columnwidth]{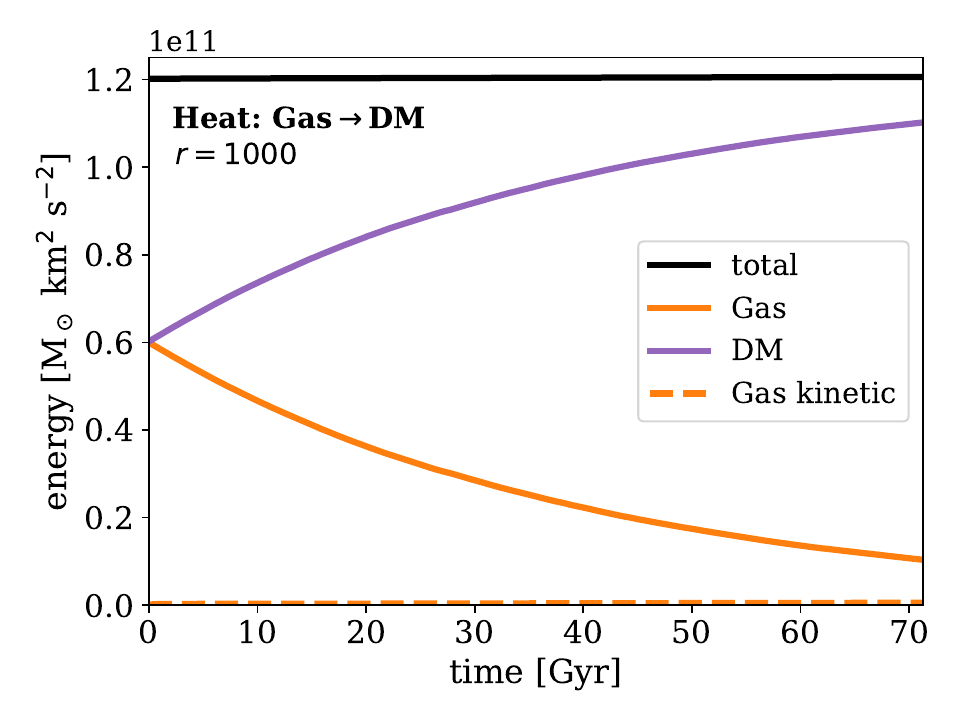}
    \caption{Same as in Fig.~\ref{fig:results_heat_dm2gas}, but for a heat conduction problem with an unequal mass ratio evolved using SPH.}
    \label{fig:results_heat_gas2dm_unequal}
\end{figure}

In Fig.~\ref{fig:results_heat_gas2dm_unequal}, the evolution of the different energy components is shown.
As expected the energy of the DM particles increases over the course of the simulation while the internal energy of the baryons is decreasing.
\begin{figure}
    \centering
    \includegraphics[width=\columnwidth]{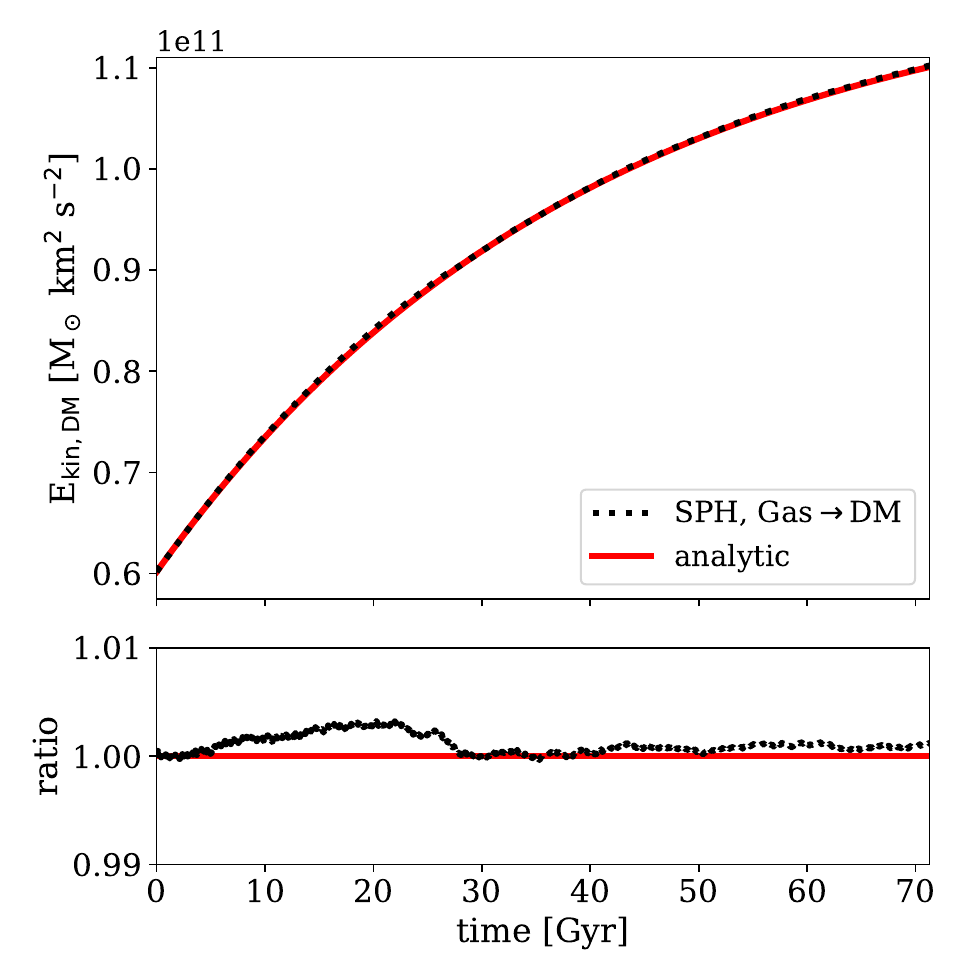}
    \caption{Kinetic energy of the DM particles as a function of time. The simulated (black) time evolution for the heat conduction problem with a mass ratio of $r=1000$ is compared to the exact solution (red) from Eqs.~\eqref{eq:heat_conduct_exact_solution} and ~\eqref{eq:heat_conduct_kappa}. The upper panel gives the absolute values and the lower panel displays the ratios to the exact solution.}
\label{fig:results_heat_gas2dm_unequal_dmkineticenergy}
\end{figure}
A detailed comparison of the simulation results to the exact solution given by Eqs.~\eqref{eq:heat_conduct_exact_solution} and ~\eqref{eq:heat_conduct_kappa} follows in Fig.~\ref{fig:results_heat_gas2dm_unequal_dmkineticenergy}. Here, we can see that the simulation results agree well with the analytic description of the test problem. Thus we can conclude that we are able to simulate fairly unequal mass ratios.

However, although these results are promising, we have to note that due to the stochastic nature of the simulation scheme, some of the baryonic particles could reach internal energies close to zero way before the average internal energy of the baryons has reached similar values. This poses a limitation on what can be simulated, it makes the set-up prone to the negative internal energy problem discussed in Sect.~\ref{sec:negative_energy_problem} (see also Sect.~\ref{sec:viscosity_and_heat_conduction}). In practice, this problem is prevented by the baryons being subject to a strong enough heat conduction (among the baryons).
If we would run the test without artificial heat conduction, it would not have been possible to evolve the system as far as shown in Figs.~\ref{fig:results_heat_gas2dm_unequal} and~\ref{fig:results_heat_gas2dm_unequal_dmkineticenergy}.

\subsection{Isotropic scattering} \label{sec:test_problems_isotropic}
We also run a simulation with an isotropic cross-section (Eq.~\eqref{eq:cross-section_iso}) and equal physical particle masses ($r=1$). Following the explanations in Sect.~\ref{sec:negative_energy_problem}, we choose the numerical baryonic particle mass to be larger than the numerical DM particle mass to avoid non-positive internal energies. For this test, we set the baryonic particle mass to about ten times the DM mass. We simulate a total cross-section per DM particle mass of $\sigma / m_\chi = 20 \, \mathrm{cm}^2 \, \mathrm{g}^{-1}$, this corresponds to a momentum transfer cross-section of $\sigma_\mathrm{T} / m_\chi = 10 \, \mathrm{cm}^2 \, \mathrm{g}^{-1}$, which is relevant for the heat exchange between the DM and baryons. Furthermore, we assume $f_\mathrm{bary} = 1$. The number of DM particles is $N_\mathrm{DM} = 10^5$ and the number of baryonic particles is $N_\mathrm{bary} = 9261$, their masses are $m_\mathrm{DM} = 10^{5} \, \mathrm{M_\odot}$ and $m_\mathrm{bary} = 1.0798 \times 10^{6} \, \mathrm{M_\odot}$.
The corresponding matter densities are $\rho_\mathrm{DM} = \rho_\mathrm{bary}= 10^7 \, \mathrm{M_\odot} \, \mathrm{kpc}^{-3}$ and the initial energy densities are the same as in Sect.~\ref{sec:test_problems_heat_conduct} ($w_\mathrm{DM} = 6 \times 10^{7} \, \mathrm{M_\odot} \, \mathrm{km}^2 \, \mathrm{s}^{-2} \, \mathrm{kpc}^{-3}$ and $w_\mathrm{bary} = 6 \times 10^6 \,\mathrm{M_\odot} \, \mathrm{km}^2 \, \mathrm{s}^{-2} \, \mathrm{kpc}^{-3}$). This implies a one-dimensional DM velocity dispersion of $\nu = 2 \, \mathrm{km} \, \mathrm{s}^{-1}$.
We employ the same neighbour numbers as before and evolve the system at a fixed time step $\Delta t = 0.024 \, \mathrm{Gyr}$.

\begin{figure}
    \centering
    \includegraphics[width=\columnwidth]{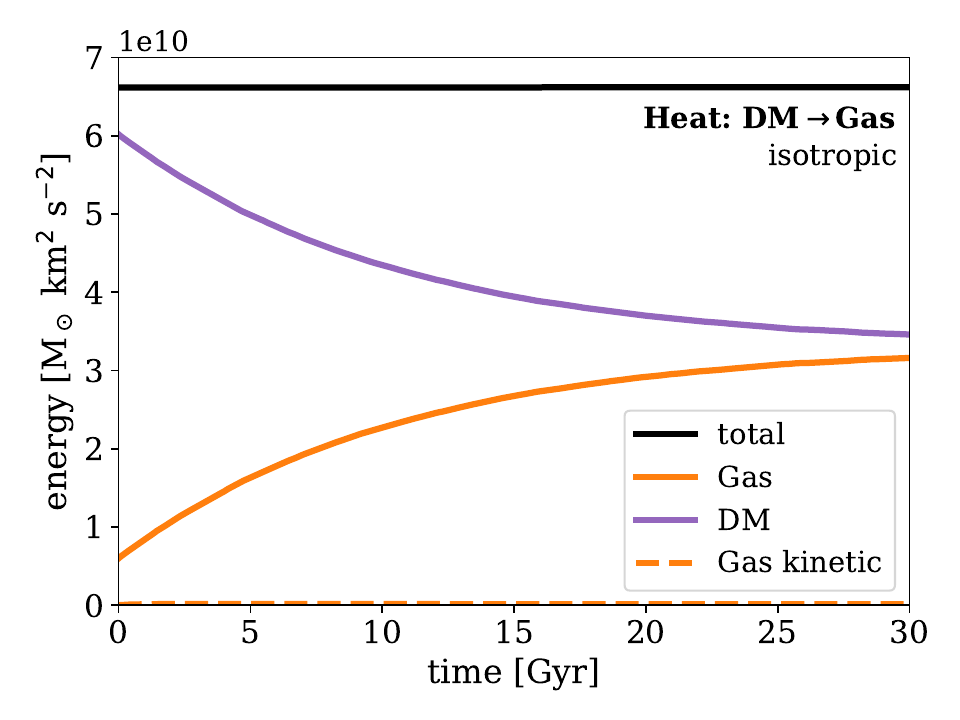}
    \caption{Energy evolution for isotropic scattering with equal particle masses. The different energy components for DM and baryons are shown as a function of time. This is the same test problem as shown in Fig.~\ref{fig:results_heat_dm2gas} for small-angle scattering.}
    \label{fig:results_heat_dm2gas_isotropic}
\end{figure}

The simulation results are shown in Fig.~\ref{fig:results_heat_dm2gas_isotropic}.
Overall they look promising as we do not find a drastic increase in total energy, it is conserved up to $0.1\%$.
Moreover, the kinetic energy of baryons stays also rather low and the baryonic and DM components evolve smoothly towards the equilibrium state.

\begin{figure}
    \centering
    \includegraphics[width=\columnwidth]{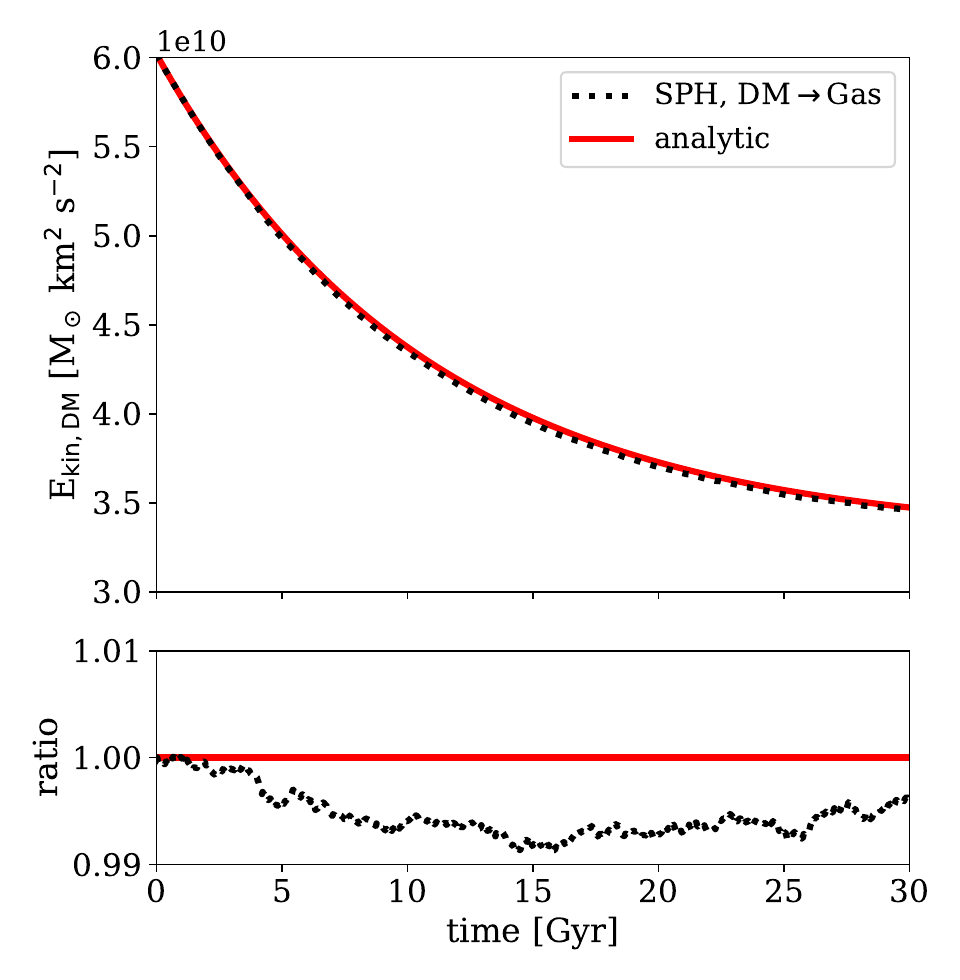}
    \caption{Time evolution of the kinetic energy of the DM. The simulation results (black) for the heat conduction problem with isotropic scattering are compared to the exact solution (red) from Eqs.~\eqref{eq:heat_conduct_exact_solution} and ~\eqref{eq:heat_conduct_kappa}.
    The upper panel gives the absolute values, and the lower panel displays the ratios to the exact solution.}
    \label{fig:results_heat_dm2gas_isotropic_dmkineticenergy}
\end{figure}

In Fig.~\ref{fig:results_heat_dm2gas_isotropic_dmkineticenergy}, we display the kinetic energy of the DM as a function of time and compare it to the exact solution from Eqs.~\eqref{eq:heat_conduct_exact_solution} and ~\eqref{eq:heat_conduct_kappa}. As we can see, the simulation results agree well with the analytic description of test problem.
Hence, we can conclude that we are not only able to simulate small-angle scattering but large-angle scattering as well, in this case, an isotropic cross-section.

\section{Halo formation} \label{sec:halo_formation}

In this section, we study the collapse of an overdensity and the formation of the halo under the influence of DM-baryon interactions. We begin with a description of our simulation set-up and subsequently show and discuss our results.
This time we only use SPH as it gave better results for the test problems of the previous section.

\subsection{Simulation set-up}
To simulate the formation of a $M \approx 10^{12} \,\mathrm{M_\odot}$ halo from the collapse of an overdensity we employ an idealized set-up. For the ICs, we sample only a single spherically symmetric overdensity at $z_i=1000$. We compute the density distribution following \cite{Lu_2006} \citep[see also Sect.~3.3 by][]{Nadler_2017}.
Specifically we employed the universal form for halo mass accretion histories as found by \cite{Wechsler_2002} from a fit to simulations,
\begin{equation}
    M(z) = M_0 \, \exp\left[\frac{-S}{1+z_c} \left(\frac{1+z}{1+z_f} -1\right)\right] \,.
\end{equation}
At redshift $z$ the virial mass is given by $M_0$, while redshift $z_c$ characterizes the point in time at which the mass accretion rate $\mathrm{d}(\log M )/\mathrm{d}(\log a)$ falls below a critical value $S = 2$.
The linear overdensity $\delta_i$ at redshift $z_i$ for a perturbation of mass $M$ that collapses at $z$ is
\begin{equation}
    \delta_i(M) = 1.686 \frac{D(z_i)}{D(z(M))} \,.
\end{equation}
For the linear growth factor $D(z)$ we use the fitting formula by \cite{Carroll_1992},
\begin{equation}
    D(z) = \frac{g(z)}{1+z} \,,
\end{equation}

\begin{equation}
    g(z) \approx \frac{5}{2} \frac{\Omega_\mathrm{M}(z) }{ \Omega^{4/7}_\mathrm{M}(z) - \Omega_\mathrm{\Lambda}(z) + \left[ 1 + \frac{\Omega_\mathrm{M}(z)}{2} \right] \left[ 1 + \frac{\Omega_\mathrm{\Lambda}(z)}{70} \right]} \,.
\end{equation}
The cosmological parameters $\Omega_\mathrm{M}(z)$ and $\Omega_\mathrm{\Lambda}(z)$ denote the density parameter of non-relativistic matter and of the cosmological constant respectively at redshift $z$.
For the mass, $M$, enclosed within a radius $r_i$ at redshift $z_i$ we use
\begin{equation} \label{eq:radius_encmass}
    r_i(M) = \left\{\frac{3 M}{4\uppi\overline{\rho}(z_i)[1+\delta_i(M)]}\right\}^{1/3} \,,
\end{equation}
where $\overline{\rho}(z_i) = \rho_\mathrm{crit,0} \, \Omega_\mathrm{M}(0)\, (1+z_i)^3$, with the critical density $\rho_\mathrm{crit,0}$ at $z=0$.
We sample the spherical overdensity by creating equal mass bins with radii according to Eq.~\eqref{eq:radius_encmass}. Next, we approximate the density within the bins by a power law and enforce continuity. The positions are sampled in a Monte Carlo fashion using direct sampling and are rearranged on spherical shells to reduce density fluctuations.
For the simulations in this section we use $z_i = 200$, $z_c = 3.0$, $z_f = 0.0$, and $M_0 = 10^{12} \, \mathrm{M_\odot}$ to generate the ICs.

We embed the overdensity in a cubic box with a comoving side length of $l_\mathrm{box} = 3698.4 \, \mathrm{ckpc}$. The volume around the overdensity is filled with a constant density $\rho(r_i(M_0))$.
The positions of the corresponding particles are sampled employing a Monte Carlo approach with rejection sampling.
In total the ICs contain $N_\mathrm{bary} = 2.5 \times 10^5$ and $N_\mathrm{DM} = 2.5 \times 10^6$ particles.
Initially, the velocity of all the numerical particles is set to zero, which implies that DM is cold. The temperature of the SPH particles is set to $T = 547.73 \, \mathrm{K}$ assuming a mean molecular weight of $\mu = 1.2 \, \mathrm{m_p}$ for a neutral gas with primordial abundances of hydrogen and helium.
Adopting the Planck 2018 results \citep{Planck_2020}, we employ the following cosmological parameters: $H_0 = 67.66 \, \mathrm{km} \mathrm{s}^{-1} \, \mathrm{Mpc}^{-1}$, $\Omega_\mathrm{M0} = 0.3106$, $\Omega_\mathrm{\Lambda0} = 0.6894$, $\Omega_\mathrm{B0} = 0.0489$.

A comoving softening length of $\epsilon = 2.96 \,\mathrm{ckpc}$ is in place and we use a fixed time step for all particles to achieve high accuracy.
We note that the time stepping is done in $\eta = \log(a)$, with $a$ being the scale factor. This implies that the time steps in physical units are not equal, but $\Delta \eta$ is constant. However, given that the required size of $\Delta \eta$ decreases substantially over the course of the simulation, we decrease the time step a few times over the course of the simulation for all particles when required.
Additionally, we use artificial viscosity and artificial heat conduction as in the previous section for the SPH simulations. 
We do not expect that these terms have a large global effect, as they are implemented time and spatial dependent to act against local discontinuities only, e.g.\ the artificial heat conduction does not lead to a coherent heat flow as the gravitational forces are taken into account \citep[the implementation has been described by][]{Beck_2016}.
Furthermore, we employ our default neighbour numbers: $N_\mathrm{ngb, SPH} = 230$, $N_\mathrm{ngb, DM} = 64$, and $N_\mathrm{idm} = 384$. The simulations are executed using the MPI parallelisation of the IDM scheme.

We simulate the collapse of the overdensity with collisionless DM and a simple IDM model with a velocity-independent forward-dominated cross-section (Eq.~\eqref{eq:cross-section_fwd}) and assume $r=1$ and $f_\mathrm{bary} = 1$. We chose the cross-section relative to the CMB constraints by \cite{Boddy_2018a}.
It corresponds to a cross-section of $\sigma_\mathrm{T}/m_\chi \approx 0.1 \, \mathrm{cm}^2 \, \mathrm{g}^{-1}$ for velocity-independent equal-mass scattering. The cross-sections we simulate are $\sigma_\mathrm{T}/m_\chi \in \{0.1, 1.0, 10.0\} \, (\mathrm{cm}^2 \, \mathrm{g}^{-1})$

In contrast to the tests of the previous section, we use here for our simulations comoving integration in an expanding space, which we tested (described in Appendix~\ref{sec:como_test}).
Moreover, we note that the local kernel size rescaling as described in Sect.~\ref{sec:rescaling_kernel_size} is very helpful to speed up these simulations.

\subsection{Results}

\begin{figure}
    \centering
    \includegraphics[width=\columnwidth]{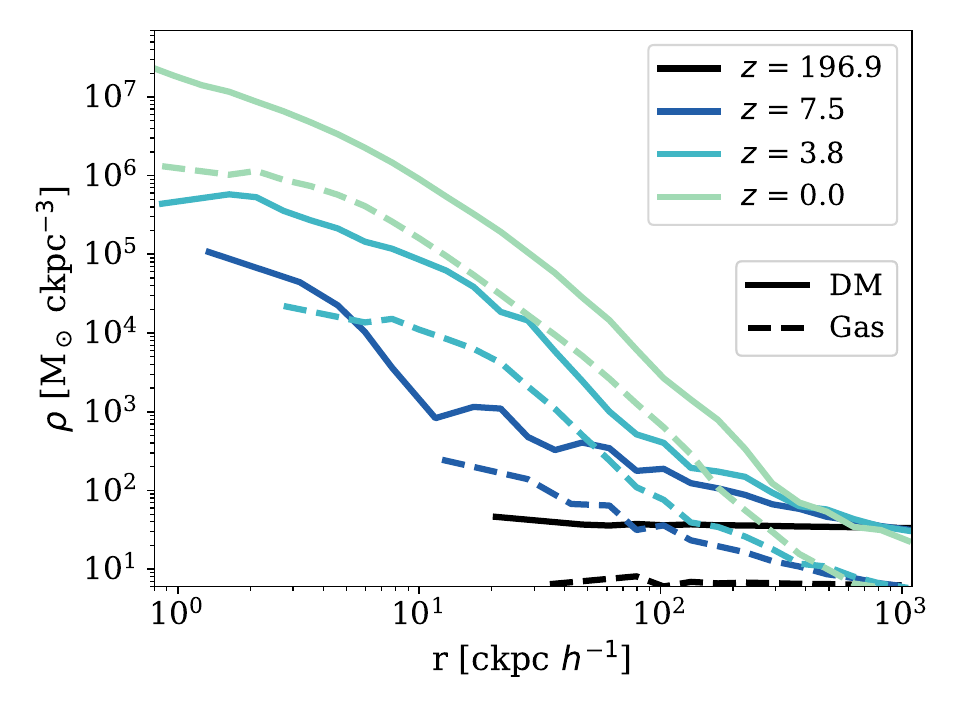}
    \caption{Density profile for a collapsing overdensity in the case of collisionless DM. The density is shown as a function of radius at several redshifts for DM (solid) and baryons (dashed). The density is not displayed for small radii where the particle number is too low to obtain a reliable value.}
    \label{fig:density_profile_cdm}
\end{figure}

\begin{figure}
    \centering
    \includegraphics[width=\columnwidth]{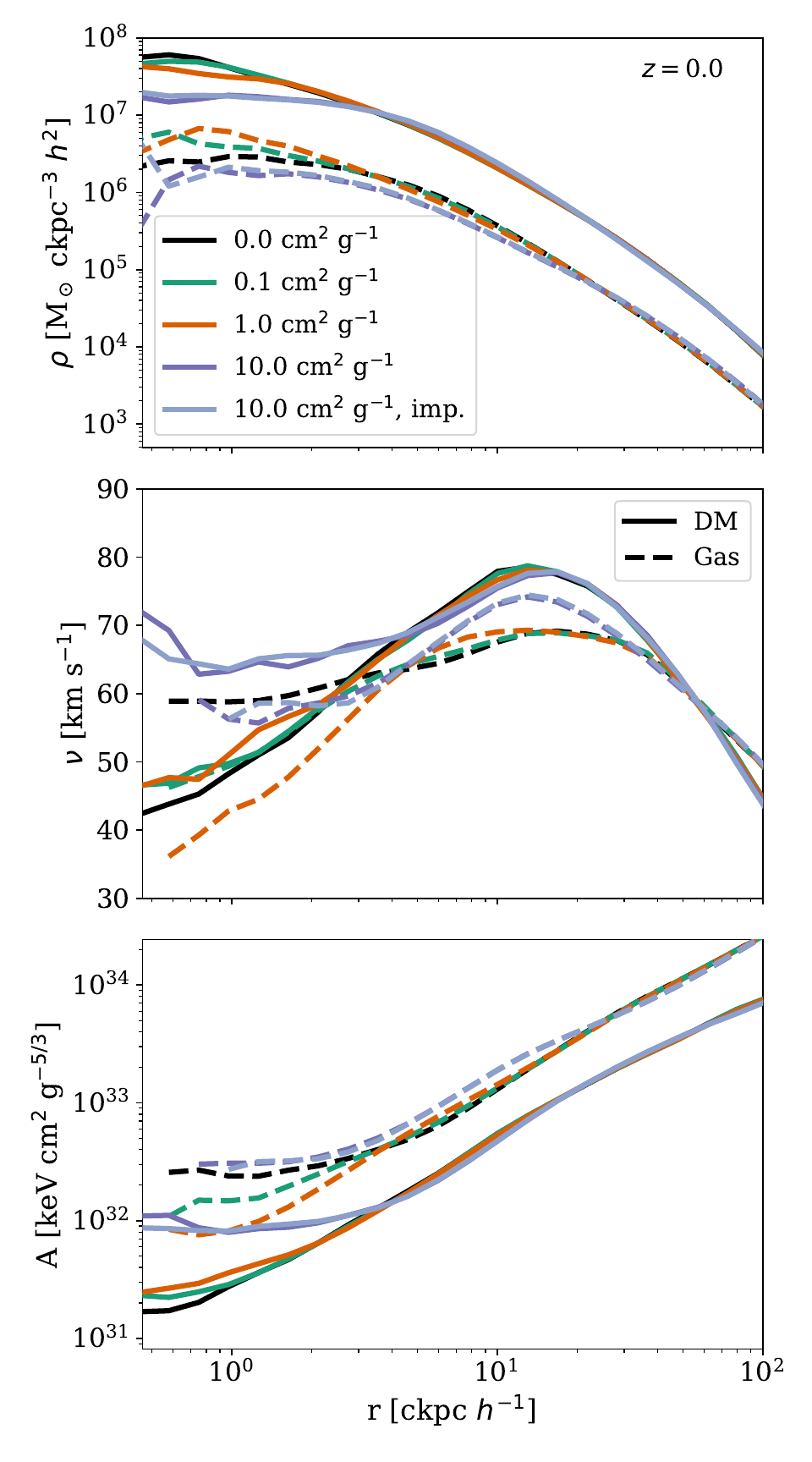}
    \caption{Density, velocity dispersion, and entropy profile of DM and baryons for CDM and IDM. We show the density as a function of radius (upper panel) at a redshift of $z=0$. The velocity dispersion, including kinetic and internal energy but excluding radial bulk motion, is displayed in the middle panel. The results for collisionless DM are shown in black. In the lower panel, the entropic function (Eq.~\eqref{eq:entropic_function}) is displayed. The solid lines are for the DM and the dashed ones are for the baryons. Furthermore, the coloured curves are for IDM simulations employing different cross-sections. In addition, the results for a simulation that makes use of the scheme for improved energy conservation as explained in Appendix~\ref{sec:improved_energy_conservation} are given too.
    }
    \label{fig:density_profile_idm}
\end{figure}

In the following, we present the results of our simulations of a collapsing overdensity. To analyse the simulations, we make use of the peak finding algorithm based on the gravitational potential presented by \cite{Fischer_2021b}. It gives us the centre with respect to which we compute other quantities such as the density profiles.

Figure~\ref{fig:density_profile_cdm} gives the comoving matter density for the DM and the baryons as a function of comoving radius. The collisionless case is shown.
Overall the comoving density is increasing for the DM and baryons with time as the overdensity collapses under gravity.
The two components evolve very similarly, except that the density of the baryons is not growing as much as for the DM at the halos centre, because of the thermal pressure.
We note that we did not include gas cooling or star formation. This allows us to compare CDM and IDM runs better and directly infer the impact of the DM-baryon scattering without being affected by additional physical processes as we do next. However, this limits the physical fidelity at the same time.

At high redshifts, the collisionless and interacting DM models behave very similarly and it is only at later times that the DM-baryon interactions lead to lower DM densities, compared to the collisionless case. As shown in the upper panel of Fig.~\ref{fig:density_profile_idm} for $z=0$, this can be understood on the basis of two mechanisms.
Firstly, at small radii, where the baryons in the absence of interactions are hotter than the DM, the scattering leads to a flow of heat from the baryons to the DM. Secondly, the velocity dispersion gradient of the DM and the temperature gradient of the baryons are positive at small radii, i.e.\ velocity dispersion and temperature increase with radius (lower panel).\footnote{The velocity dispersion in Fig.~\ref{fig:density_profile_idm} is computed from the kinetic and internal energy with the energy of the radial bulk motion being subtracted. We note, that the bulk motion is computed separately for DM and baryons, i.e.\ the two components may differ in their bulk motion.} Similarly to SIDM halos the interactions can give rise to heat transport following the velocity dispersion/temperature gradient. Given that the gradient is positive, heat is transported inward, which contributes to the formation of a density core.

Interestingly, we find for the cross-section of $\sigma_\mathrm{T} / m_\chi = 1.0 \, \mathrm{cm}^2\,\mathrm{g}^{-1}$ that the baryon density at small radii increases (upper panel of Fig.~\ref{fig:density_profile_idm}).
This can be understood with the heat exchange between DM and baryons at these radii. 
The interactions effectively cool the baryons (middle panel of Fig.~\ref{fig:density_profile_idm}), which makes them contract and leads to a higher baryon density in the centre.
For the larger cross-section ($\sigma_\mathrm{T} / m_\chi = 10.0 \, \mathrm{cm}^2\,\mathrm{g}^{-1}$) we do not find this higher baryon density, because much more heat from larger radii is transported inwards leading to a higher baryon temperature compared to the $\sigma_\mathrm{T} / m_\chi = 1.0 \, \mathrm{cm}^2\,\mathrm{g}^{-1}$ case.
The smallest cross-section of $\sigma_\mathrm{T} / m_\chi = 0.1 \, \mathrm{cm}^2\,\mathrm{g}^{-1}$ has only a small impact. It decreases the DM density and increases the baryonic density due to effectively cooling the baryons (Fig.~\ref{fig:density_profile_idm}).

In summary, we find that the interplay of local heat exchange between the two components and heat inflow from larger radii leads to a non-monotonic behaviour of the central baryon density and temperature with cross-section. In contrast, the central DM density and its velocity dispersion behave monotonically with cross-section for the range we are studying here.

We want to note that we additionally simulated the set-up with $\sigma_\mathrm{T} / m_\chi = 10.0 \, \mathrm{cm}^2\,\mathrm{g}^{-1}$, using the scheme for improved energy conservation (Appendix~\ref{sec:improved_energy_conservation}). As visible in Fig.~\ref{fig:density_profile_idm}, the default scheme does not deviate significantly from the improved scheme for our set-up of the collapsing overdensity. In conclusion, the default scheme is accurate enough for these simulations.

In addition to density and velocity dispersion, we also computed the entropic function,
\begin{equation} \label{eq:entropic_function}
    A = (\gamma -1) \, u \, \rho^{1-\gamma} \,, 
\end{equation}
which is closely related to entropy. It is shown in the lower panel of Fig.~\ref{fig:density_profile_idm}. Here, we also find the non-monotonic impact of the DM-baryon interactions with cross-section on the baryon density.

\begin{figure}
    \centering
    \includegraphics[width=\columnwidth]{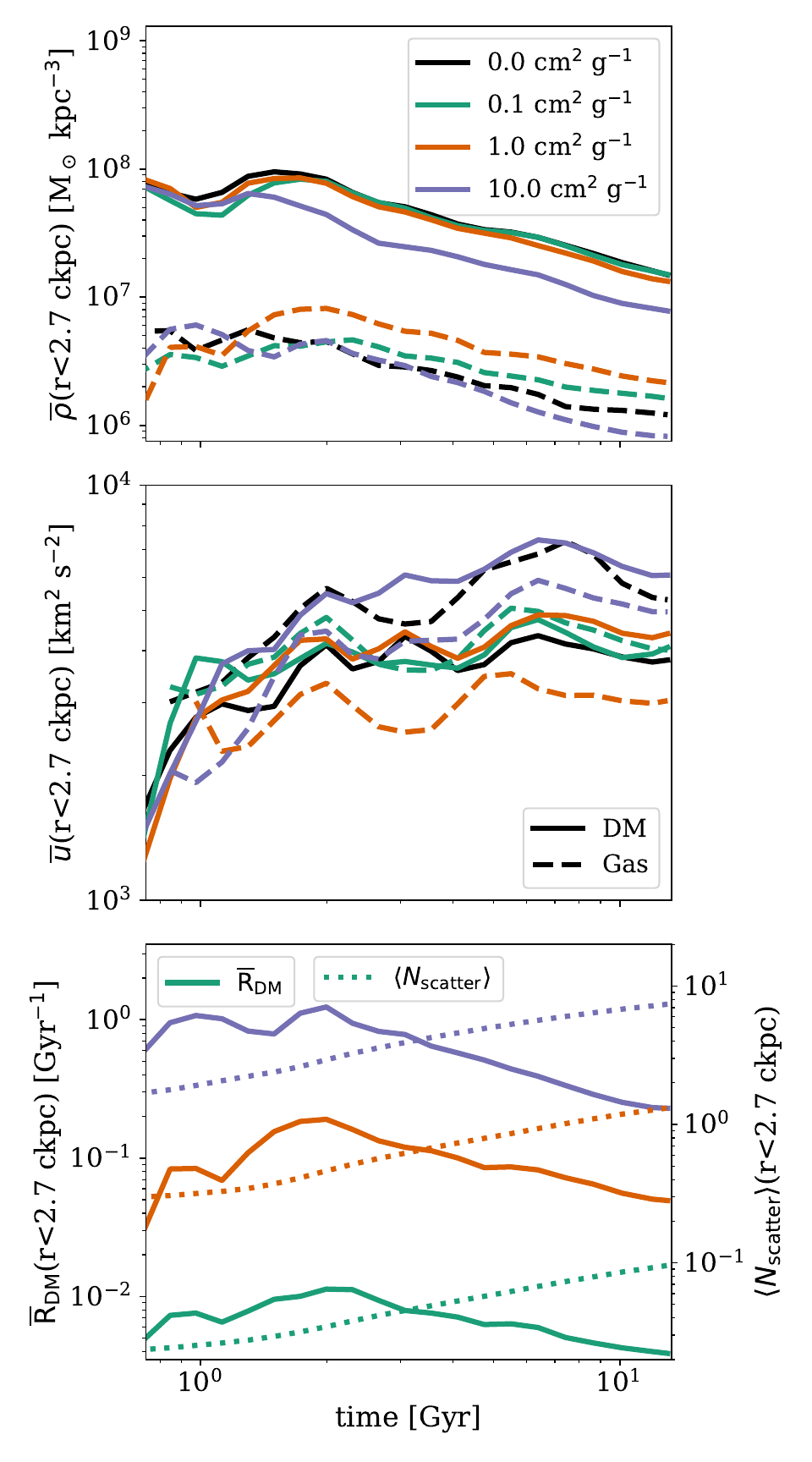}
    \caption{Central density, specific energy and DM interaction rate as a function of scale factor. For the simulated DM models we show the evolution of different quantities. The central density (upper panel) and specific energy (middle panel) within a radius of $r = 2.7 \, \mathrm{ckpc}$ are shown for the DM and baryonic components. The lower panel gives the average DM interaction rate within the same radius according to Eq.~\eqref{eq:interaction_rate} assuming an isotropic cross-section matched to the simulated forward dominated one with the momentum transfer cross-section ($\sigma|_\mathrm{iso} = 2 \, \sigma_\mathrm{T}|_\mathrm{iso}$).}
    \label{fig:results_interaction_rate}
\end{figure}

Besides the comoving density profile, we also measure the average density and specific energy within a comoving radius of $r = 2.7 \, \mathrm{ckpc}$, and compute the corresponding interaction rate of the DM particles. The results for our simulations are shown in Fig.~\ref{fig:results_interaction_rate}.
We note, that the shown specific energy and densities are not comoving but physical. For the specific energy, $u$, we compute the sum of the kinetic and internal energy divided by mass. This specific energy is then used to compute the interaction rate, $R_\mathrm{DM}$, of the DM particles shown in the lower panel. For a velocity-independent cross-section, it is given by,
\begin{equation} \label{eq:interaction_rate}
    R_\mathrm{DM} = 4 \frac{\sigma}{m_\chi} \, \rho_\mathrm{bary} \, \sqrt{\frac{u_\mathrm{DM} + u_\mathrm{bary}}{3 \uppi}} \, .
\end{equation}
We note for the simulated forward-dominated model (Eq.~\eqref{eq:cross-section_fwd}) the total cross-section $\sigma$ does not have a finite value. Instead, we assume that the impact of the DM-baryon interactions could be matched to isotropic scattering by employing the momentum transfer cross-section (Eq.~\eqref{eq:transfer_cross-section}). For isotropic scattering the two cross-sections are related to each other through $\sigma|_\mathrm{iso} = 2 \, \sigma_\mathrm{T}|_\mathrm{iso}$.

In the upper panel of Fig.~\ref{fig:results_interaction_rate} we can see that the central density for DM and baryons decreases with time. At the same time, the specific energy (middle panel) increases and reaches very similar values for DM and baryons.
We note that we only show the late part of the evolution where the DM-baryon interactions have a significant impact on the density profile. At the earlier times, which are not shown, CDM and IDM behave almost the same.
In principle, the DM interaction rate is fairly high at early times but decreases very quickly and stays for almost all of the cosmic evolution much lower. Given that $R_\mathrm{DM}$ decreases very quickly, the high values at early times have very little impact.
In the lower panel of Fig.~\ref{fig:results_interaction_rate}, we show only the interaction rate at later times but also give the expected number of scattering events a particle has undergone since $z=200$.
For the collisionless case, the interaction rate is always zero.

Overall, our simulations demonstrate that we can model the impact of DM-baryon interaction on physically interesting systems. The most noticeable effects result from cross-sections that are large compared to the CMB bounds by, for instance, a factor of 100 ($\sigma_\mathrm{T}/m_\chi = 10.0 \, \mathrm{cm}^2 \, \mathrm{g}^{-1}$). While such a large cross-section can lead to a sizeable density core, a cross-section of $\sigma_\mathrm{T}/m_\chi = 0.1 \,\mathrm{cm}^2 \, \mathrm{g}^{-1}$, which roughly corresponds to the CMB bound, leads to a much more subtle effect, but could still increase the central baryon density by a factor of two. To more fully unpack the effects of DM--baryon scattering at low redshifts, several steps beyond the scope of our work are necessary. First, we only studied one system with a mass of $10^{12} \, \mathrm{M}_\odot$, but this could be extended to other systems as well. 
Second, we neglected various baryonic processes. The inclusion of gas cooling would lead to higher densities in the centre and thus increase the interaction rate (at least at a fixed relative DM--baryon velocity), whereas, for example, supernova feedback could have the opposite effect. However, when taking the velocity dependence into account, the picture could become more complicated. This leads us to the third point, we only considered a velocity-independent cross-section with scattering partners having equal masses. This could be extended to a velocity-dependent cross-section with unequal-mass scattering.
Fourth, we assumed that there is no relative motion between the DM and the baryons for the initial conditions of our simulations (at $z=200$). However, there could be a significant relative velocity \citep[e.g.][]{Tseliakhovich_2010, Dvorkin_2014}. If this is the case, the impact of the DM-baryon interactions might be stronger than what one would expect based on our simulations.
IDM simulations like ours can be a valuable tool for deriving bounds on the interaction cross-section, particularly for the impact of DM-baryon scattering at low redshift. Finally, we assumed ICs (in the form of an initial density perturbation for halo collapse) consistent with CDM, whereas IDM can significantly suppress the linear matter power spectrum and thus delay (or even prevent) the formation of halos below a critical mass that depends on the IDM mass and cross-section~\citep{Nadler_2019,Nadler_2025a}. We leave a self-consistent treatment of halo formation in IDM to future work.

\section{Discussion} \label{sec:discussion}
In this section, we discuss various aspects of our numerical scheme for IDM and the simulations we ran. In particular, we highlight the remaining challenges and discuss potential next steps for investigating the numerics and physics of DM-baryon interactions.

In Sect.~\ref{sec:test_problems}, we tested our IDM implementation with two schemes for hydrodynamics, namely SPH and MFM. We found that those schemes show significant differences for the tests we conducted. These differences stem from how the fluid equations are solved. For example, to conduct an SPH simulation we need artificial viscosity while for MFM this is not the case. Moreover, we found differences in the capability of conserving the total energy, which is related to the time integration, as we discuss in Sect.~\ref{sec:energy_conservation} and Appendix~\ref{sec:improved_energy_conservation}. There exist further numerical schemes to solve the fluid equations, for example, there are a variety of advanced SPH schemes \citep[for a review see][]{Zhang_2022}. It is left for future work to understand how those schemes can be best combined with IDM and to investigate which of them might be best suited for simulations incorporating DM-baryon interactions.

We have found that viscosity and heat conduction can play a crucial role in making the numerical scheme robust. In particular, we discuss the fact that a negative internal energy can arise if the interactions are strong compared to viscosity and heat conduction (see Sect.~\ref{sec:negative_energy_problem}). Depending on the numerical scheme for hydrodynamics this can make the use of artificial viscosity and artificial heat conduction crucial. 
Ideally, this is done in a fashion that acts against local discontinuities only and does not alter the properties on larger scales of the simulated system. In this line, we used the implementation described by \cite{Beck_2016}.
Moreover, we found that the numerical formulation of the DM-baryon interactions introduces small-scale turbulence, which can be suppressed by viscosity, but must not harm the overall evolution of the system as turbulent pressure compensates for the missing thermal pressure. However, at the same time, this limits the interpretation of the hydrodynamical properties as we might not be able to clearly distinguish between the temperature and small-scale motion of the baryons.

Another aspect that can be crucial in making the scheme numerically robust is the choice of the numerical particle masses. In general, it is preferable to choose the masses for the baryons larger than for DM. This becomes particularly important for large scattering angles (see Sect.~\ref{sec:negative_energy_problem}). Unfortunately, this stands in contrast to the way particle masses are typically chosen. For cosmological simulations of galaxy formation, it is common to choose the mass of the baryonic particles smaller than the one of the DM particles. To achieve the mass resolution for the baryons, many more DM particles would be needed making the simulation more expensive. 
The baryonic particle mass should also not be chosen too large because gravitational interactions can then lead to artificial mass segregation \citep[e.g.][]{Ludlow_2019, Ludlow_2023}. 

We found that numerically ensuring energy conservation is more challenging for IDM than for SIDM simulations. These issues are related to the fact that for IDM we also change the internal energies of the baryonic particles and not only the velocities, potentially interfering with the time integration. At least for SPH, we have been able to demonstrate that it is possible to improve energy conservation by modifying the time integration scheme with a moderate increase in computational costs (see Appendix~\ref{sec:improved_energy_conservation}). However, in the case of MFM, it is left for future work to investigate how improvements in energy conservation could be achieved. An important aspect in this context is the use of variable time steps. So far, we have restricted ourselves to fixed time steps because we found that variable time steps can lead to a somewhat lower accuracy. It remains an open question how the time integration scheme can be improved when using variable time steps. This is left for future work, but it constitutes an important aspect given that it would allow for drastically reduced computational costs for typical astrophysical systems while maintaining high accuracy.

Overall, the aspects we mention in this paper indicate that there are possibilities and parameters that can be tuned to achieve a high accuracy of the IDM simulations. However, at the same time, they come with an increase in computational costs. One may have to find the best trade-off in respect of the specific problem at hand. There might not be simple guidance given the non-trivial interplay of the different aspects such as the size of the time steps, the numerical particle mass, the interaction number $N_\mathrm{idm}$, viscosity, heat conduction, and the size of the scattering angles.

We note that the inclusion of additional processes could interfere with the numerical scheme and degrade its numerical properties, leading to larger numerical errors.
We expect that simulating gravity along with the DM-baryon interactions and hydrodynamics is unproblematic as long as the gravitational accelerations are computed based on the positions only, in contrast to higher-order schemes that employ the particle velocities as well.
This is in line with the finding by \cite{Fischer_2024b} that the inclusion of their scheme for DM self-interactions does not hurt the vanishing drift in total energy for a symplectic and time-symmetric leap-frog integrator.
However, problems that are even present without IDM are not expected to vanish when it is included. This can be seen analogously to the difficulties found in simulations of gravothermally collapsing SIDM halos. The main problems that show up are already present in a gravity-only simulation, but are amplified in the collapsing halos due to the high concentration of these objects \citep{Fischer_2024b}. 

On a similar note, the simulations of a collapsing overdensity presented in Sect.~\ref{sec:halo_formation} are challenging. They are resolved by $2.75 \times 10^{6}$ particles. Although one may naively assume that this resolution is sufficiently high, we want to note that the identified minimum in the gravitational potential, which we use as the centre, fluctuates. Hence, inferred quantities suffer from this noise. However, simply using the centre as specified when generating the ICs does not help. This problem stems from artificial fragmentation and improves when increasing the resolution. That artificial fragmentation can be problematic in $N$-body simulations is well known and can in principle be resolved by using a different numerical scheme \citep[e.g.][]{Hahn_2013, Hahn_2015}.

In this paper, we consider only the case where DM particles scatter off one species of baryonic particles. However, an interesting case is scatterings of DM with the nuclei of Hydrogen and Helium, the two most abundant elements in the Universe. It would be straightforward to extend our numerical scheme to this case. For each species, the interactions can be computed following our description and employing the correct parameters, such as the mass ratio $r$ and the mass fraction of the baryonic species $f_\mathrm{bary}$. The scattering routine that we implemented would then simply be executed twice per time step, for each species once using the corresponding parameters. Besides, the velocity distribution (see Eq.~\eqref{eq:maxwell}) from which the velocity of the virtual particle is drawn needs to be adjusted in the case of baryonic species with unequal masses such as Hydrogen and Helium. Moreover, each species would have its own time step criterion.

Our numerical method could be applied to various astrophysical systems, as we know it from SIDM, to constrain the DM-baryon interactions. The following systems could prove particularly interesting for IDM:\ \begin{itemize}
    \item[a)] have large difference in temperature between the DM and baryonic component:
    \item [b)] show a significant relative motion between the DM and the baryons.
\end{itemize}
Systems falling into category a) can give rise to a significant heat flux between DM and baryons due to the DM-baryon scattering. Studies in this regard have been conducted for galaxy clusters \citep[e.g.][]{Shoji_2024, Stuart_2024}. Potentially also less massive systems could be of interest.
An example for a system that falls into category b) would a dissociative galaxy cluster merger. Here the DM can move with a relatively high velocity relative to the intra-cluster medium. IDM could eventually produce small offsets between the distribution of galaxies and the DM component as it is known from SIDM \citep[e.g.][]{Robertson_2017a, Fischer_2023b, Valdarnini_2023}.

When extending the application of the IDM scheme to other physical systems and higher fidelity of modelling them, it could become relevant to include gas cooling as well as stars and black holes in the simulations and their associated processes. Those processes when happening on scales that cannot be resolved are termed `subgrid physics' and are described by effective models, for example, star formation. Depending on how these models are formulated, they could plausibly interfere with the DM-baryon interactions. As we discuss in Sect.~\ref{sec:viscosity_and_heat_conduction}, our scheme could lead to artificial small-scale turbulence and artificially large variations in the internal energy, potentially affecting subgrid models. Investigating the compatibility with various subgrid models is left to a future work.

\section{Conclusions} \label{sec:conclusion}

In this work, we present a novel method for simulating interactions between DM and baryons within the framework of the $N$-body method.
It allows us to simulate a wide range of models with different mass ratios of the interacting particles, comprising a cross-section of varying angular and velocity dependencies.
Our scheme allows for the simulation of typical astrophysical set-ups and we have demonstrated its accuracy with several test problems introduced for this purpose.
In addition, we simulated the collapse of an overdensity to study the effect of IDM on halo formation.
Furthermore, we have discussed the limitations of the scheme and directions for future development.
Our main conclusions are as follows: 
\begin{enumerate}
    \item It is possible to simulate various models of IDM self-consistently within $N$-body simulations. In particular, we are able to simulate unequal-mass scatterings more easily than in state-of-the-art SIDM schemes.
    \item Artificial viscosity and heat conduction can play an important role in ensuring numerical stability for simulations involving a large cross-section compared to the viscosity and heat conduction of the baryons.
    \item For simulations of low-viscosity systems, it might not always be possible to clearly distinguish between the turbulence and temperature of the baryons on small scales.
    \item With our simulations of a collapsing overdensity, we have demonstrated that it is possible to simulate a physically relevant system and found that a velocity-independent cross-section at the level of the CMB bounds could exert only a small dynamical impact on the collapse of a Milky Way-mass halo.
Since we made several highly simplifying assumptions, such as neglecting the gas cooling, a more sophisticated study is needed to make a definitive assessment.
    \item Moreover, we found in those simulations that the central baryon density and temperature do not respond monotonically to an increase in the cross-section, which is in contrast to the behaviour of the DM (Fig.~\ref{fig:density_profile_idm}).
\end{enumerate}
This paper constitutes only a first step towards exploring the full astrophysical phenomenology of DM-baryon interactions by providing a numerical method. Future studies that use complementary IDM simulation methods (Wen et al., in preparation) and explore the effects of these interactions on diverse astrophysical systems could help constrain the cross-section of DM-baryon interactions and shed new light on the nature of DM.

\begin{acknowledgements}
The authors thank Ludwig Daniel Schmidt for his feedback on the analytic solution to the heat conduction problem.
This work is funded by the \emph{Deutsche Forschungsgemeinschaft (DFG, German Research Foundation)} under Germany’s Excellence Strategy -- EXC-2094 ``Origins'' -- 390783311.
KD, MSF and FG acknowledge support by the COMPLEX project from the European Research Council (ERC) under the European Union’s Horizon 2020 research and innovation program grant agreement ERC-2019-AdG 882679.

Software:
NumPy \citep{NumPy},
Matplotlib \citep{Matplotlib}.
\end{acknowledgements}

\bibliographystyle{aa}
\bibliography{bib.bib}

\begin{appendix}
\section{Derivation of interaction probability and drag force} \label{sec:derivation_p_and_drag}

In this appendix, we derive the interaction probability (Eq.~\eqref{eq:probability}) for the rare scattering scheme as well as the drag force (Eq.~\eqref{eq:drag_force}) for the frequent scatterings in the case of DM-baryon interactions.
Our derivations follow closely the ones presented by \cite{Fischer_2021a} for the case of SIDM.

\subsection{Interaction probability}
First, we derived the interaction probability by starting from the probability of a DM particle scattering about a baryonic particle.
The probability, $P_\mathrm{scatter}$, depends on the total cross-section, $\sigma(v)$, which can be a function of the scattering velocity, $v$.
The DM particle is assumed to travel through a constant baryonic density $\rho_\mathrm{bary}$ for a time $t$.
In case not all baryonic particles are interacting with the DM, $f_\mathrm{bary}$ gives the mass fraction of particles taking part in the interaction.
The scattering probability is,
\begin{equation}
    P_\mathrm{scatter} = \sigma(v) \, \frac{f_\mathrm{bary} \, \rho_\mathrm{bary}}{r \, m_\chi} \, v \, t \,.
\end{equation}
We note that $m_\chi$ denotes the mass of the DM particles and r is the mass ratio between the physical baryonic particle, for instance,\ a proton, and the physical DM particle.

The scattering velocity is $v = |\mathbf{v}_i - \mathbf{v}_j|$, where one of the particles is the numerical DM particle and the other one is the virtual particle.
We can compute the expected number of scatterings $\langle N \rangle$ between DM and baryons by integrating over the DM, $\rho_\mathrm{DM}$, and baryon, $\rho_\mathrm{bary}$, density.
\begin{equation}
    \langle N \rangle = \int \frac{\rho_\mathrm{DM} \, \, f_\mathrm{bary} \, \rho_\mathrm{bary}}{r \, m_\chi^2} \, \sigma(v) \, v \, \Delta t \, \mathrm{d}V \,.
\end{equation}
The densities can be interpreted as the mass represented by the numerical particles, i.e.\ $\rho_i = m_i \, W(|\mathbf{x} - \mathbf{x}_i|,h_i)$.
Using Eq.~\eqref{eq:overlap}, we can write,
\begin{equation}
    \langle N \rangle = N_\mathrm{DM} \, N_\mathrm{bary} \, \sigma(v) \, v \, \Delta t \, \Lambda_{ij} \,.
\end{equation}
Here, $N_\mathrm{DM}$ denotes the number of physical particles represented by the numerical particles. But $N_\mathrm{bary} = f_\mathrm{bary} \, m_\mathrm{bary} / (r \, m_\chi)$ is only the number of baryonic particles that could scatter. As each scattering event involves one physical DM particle and one baryonic particle, $N_\mathrm{DM} = N_\mathrm{virt} = m_\mathrm{virt} / (r \, m_\chi)$ must hold.
With the expected number of scattering events, we can easily express the probability, $P_{ij} = \langle N \rangle / N_\mathrm{DM}$, for two numerical particles to interact.
\begin{equation} \label{eq:probability2}
    P_{ij} = \frac{\sigma(v)}{m_\chi} \, \frac{f_\mathrm{bary}}{\mu} \, m_\mathrm{DM} \, v \, \Delta t \, \Lambda_{ij} \,.
\end{equation}

\subsection{Drag force}
Second, we provide a derivation of the drag force for the DM-baryon interactions. As done by \cite{Kahlhoefer_2014}, we start with a physical DM particle travelling through a background density $\rho$.
When the DM particle interacts with a baryonic particle, it scatters in the centre-of-mass frame about an angle $\theta_\mathrm{cms}$ and changes its momentum.
We are interested in the momentum change parallel to the direction of motion,
\begin{equation}
    \Delta p_\parallel = p_\mathrm{DM} \, \left(1-\cos \theta_\mathrm{cms}\right) \,,
\end{equation}
as it gives rise to the drag force. The idea behind this is that the momentum changes perpendicular to the direction of motion average out over many scattering events, but the ones parallel to the direction of motion accumulate and effectively decelerate the particle.

Per time $\mathrm{d}t$, we have $\mathrm{d}C$ interactions.
\begin{equation}
    \mathrm{d}C = \frac{\rho}{r \, m_\chi} \, \frac{\mathrm{d}\sigma}{\mathrm{d}\Omega_\mathrm{cms}} \, v \, \mathrm{d}t \, \mathrm{d}\Omega_\mathrm{cms} \,.
\end{equation}
The number of interactions depends on the differential cross-section, $\mathrm{d}\sigma / \mathrm{d}\Omega_\mathrm{cms}$ and the relative velocity $v$.
We can express the change of parallel momentum that a physical particle experiences as
\begin{equation}
    \mathrm{d}p_\parallel = \frac{\rho}{r \, m_\chi} \, \frac{\mathrm{d}\sigma(v)}{\mathrm{d}\Omega_\mathrm{cms}} \, v \, p_\mathrm{DM} \, \mathrm{d}t \, \mathrm{d}\Omega_\mathrm{cms} \,.
\end{equation}
Here, we use the momentum of the DM particle in the centre-of-mass frame. It is given by
\begin{equation}
    p_\mathrm{DM} = m_\chi \, v \, \frac{r}{1+r} \,.
\end{equation}
After integrating over the differential cross-section one obtains the momentum transfer cross-section, $\sigma_\mathrm{T}$ \citep[see Appendix A by][]{Kahlhoefer_2014}.
This allows for the change in the parallel momentum of a physical DM particle travelling through a baryonic background density to be expressed as
\begin{equation}
    \frac{\mathrm{d}p_\parallel}{\mathrm{d}t} = \sigma_\mathrm{T}(v) \, \frac{f_\mathrm{bary} \, \rho_\mathrm{bary} \, v^2}{1+r} \,.
\end{equation}
We note that $f_\mathrm{bary} \, \rho_\mathrm{bary}$ gives the density of particles that the DM particle could interact with.
The drag force acting on the numerical DM particle follows by integrating over the number density of DM particles it represents. 
\begin{equation}
    F_\mathrm{drag} = \int \frac{\mathrm{d}p_\parallel}{\mathrm{d}t} \, \frac{\rho_\mathrm{DM}}{m_\chi} \, \mathrm{d}V \,.
\end{equation}
Using Eq.~\eqref{eq:overlap} and $\mu = m_\mathrm{virt} / m_\mathrm{bary}$, we can derive the final equation for the drag force. For the implementation we use,
\begin{equation} \label{eq:drag_force2}
    F_\mathrm{drag} = \frac{\sigma_\mathrm{T}(v)}{m_\chi} \, \frac{f_\mathrm{bary}}{\mu \, (1+r)} \, m_\mathrm{virt} \, m_\mathrm{DM} \, v^2 \, \Lambda_{ij} \,.
\end{equation}

\section{Improving energy conservation} \label{sec:improved_energy_conservation}

\begin{figure}
    \centering
    \includegraphics[width=\columnwidth]{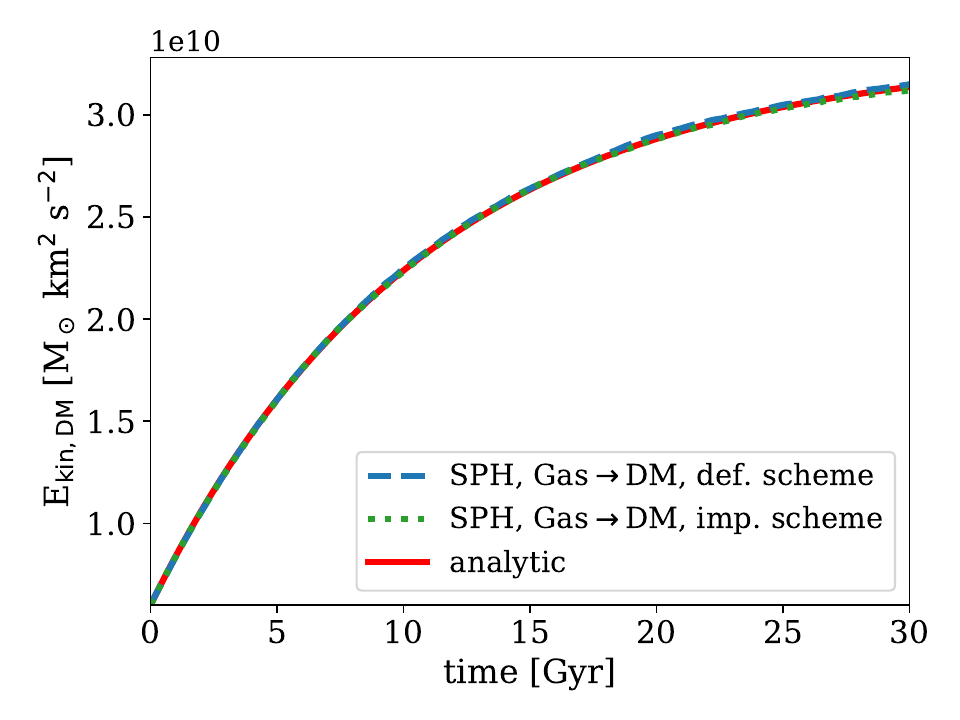}
    \caption{Kinetic energy evolution of DM for the heat conduction problem. We show the results for the test problem with heat flow from the baryons to DM for default (blue) and improved (green) implementations. In addition, the red line displays the exact solution given by Eqs.~\eqref{eq:heat_conduct_exact_solution} and~\eqref{eq:heat_conduct_kappa}.}
    \label{fig:etest_dm_kinetic}
\end{figure}

To understand how the coupling between the hydrodynamics scheme and the IDM kicks affect the energy conservation, we performed a few tests. Here, we describe how energy conservation can be improved in the case of SPH. We note, however, that based on what we found in our first tests, this does not apply to MFM in a straightforward way. In terms of energy conservation, our implementation for IDM with MFM works quite well and better compared to SPH as long as no modification of the internal energy is involved even if velocity-dependent terms such as viscosity are in place. However, when the internal energy of the gas particles is modified, SPH performs better and we are able to find a way to further improve energy conservation. Clearly, more work is needed for both MFM and SPH. To give a bit of guidance, we now explain how an improvement in the case of SPH is possible.

For improving the energy conservation, we alter the time integration and split the drift step of the KDK scheme into two half-step drifts and do the IDM kicks with the update of the internal energy in between those two half-step drifts. Previously, the DM-baryon interactions were computed between the two half-step kicks. In addition, we also recompute the hydrodynamic accelerations directly after the DM-baryon interactions and before the second half-step drift. Thus, the hydrodynamic accelerations are computed twice per time step, before the second half-step drift and before the second half-step kick, where they are usually computed only \citep[see e.g.][]{Price_2018, Groth_2023}. These modifications allow us to significantly improve energy conservation as we show below.

\begin{figure}
    \centering
    \includegraphics[width=\columnwidth]{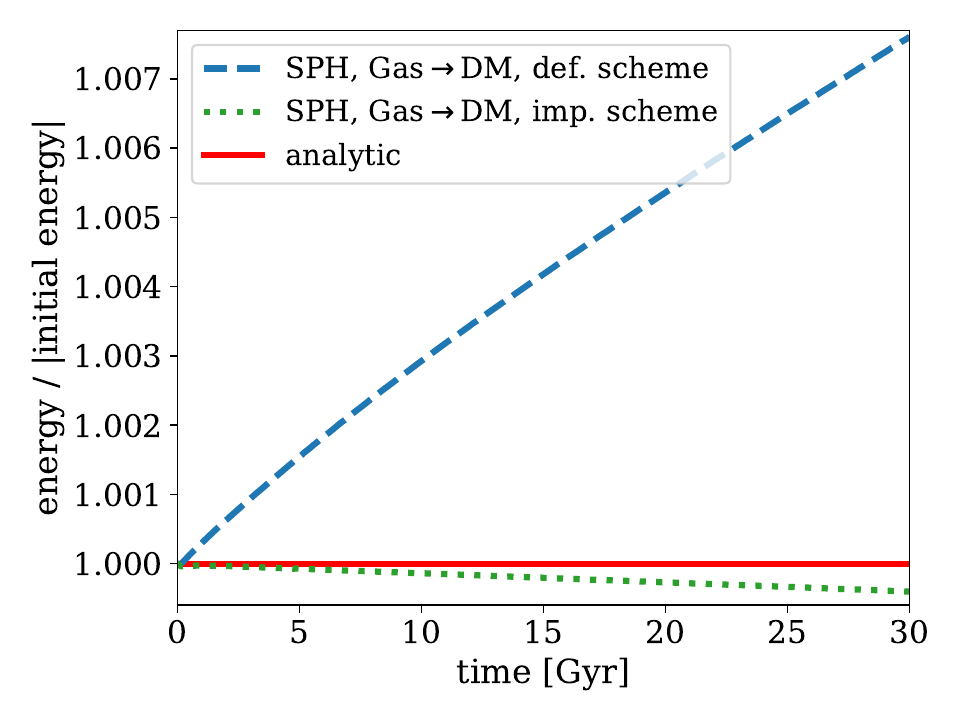}
    \caption{Energy conservation for the heat conduction problem with the default and improved implementation. From the same simulations as used for Fig.~\ref{fig:etest_dm_kinetic}, we show the energy conservation as a function of time for the default (blue) and improved (green) implementations. The expectation, i.e.\ perfect energy conservation, is indicated by the red line.}
    \label{fig:etest_energy_conservation}
\end{figure}

In Fig.~\ref{fig:etest_dm_kinetic}, we display the evolution of the kinetic energy of DM for the heat transfer problem where heat flows from the baryons to the DM. This is the same set-up used in Sect.~\ref{sec:test_problems_heat_conduct}. We show the results from our two implementations, the default one and the improved one. Both show a very similar agreement with the exact solution.
The improvement in energy conservation due to the modified time integration scheme is visible from Fig.~\ref{fig:etest_energy_conservation}. We managed to largely improve compared to the default implementation.
We note that for this test we only consider the case where all particles reside on the same fixed time step. An extension to variable time steps as well as improving the energy conservation when using MFM is left for future work.

\section{Convergence test} \label{sec:convergence}

\FloatBarrier

\begin{figure}
    \centering
    \includegraphics[width=\columnwidth]{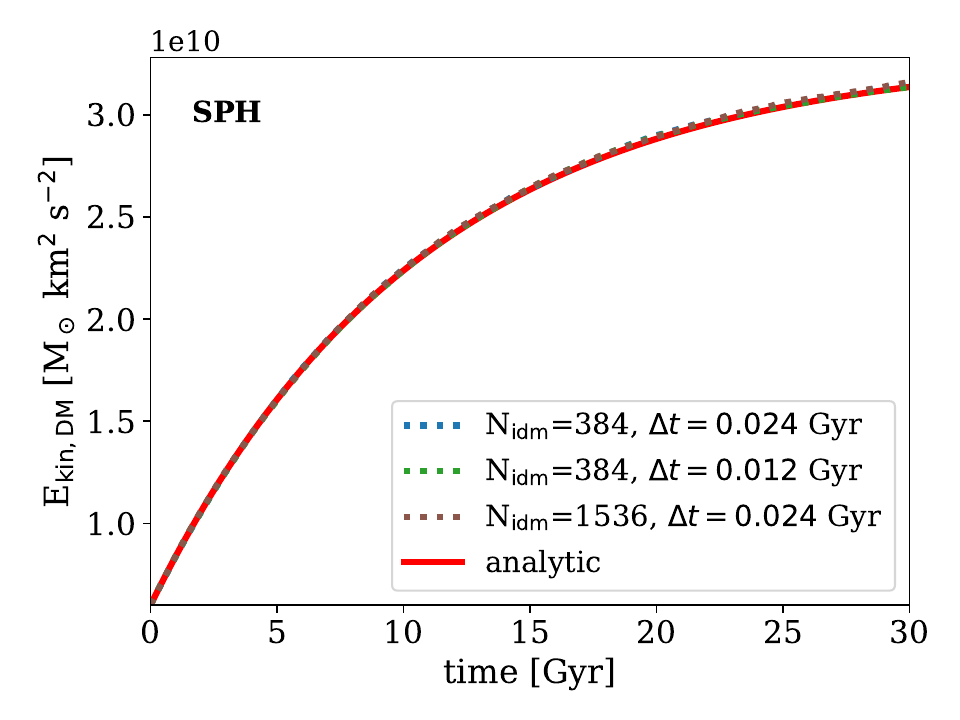}
    \caption{Test of varying the time step and the interaction number for the heat conduction problem for SPH. We show the kinetic energy of the DM particles for the problem with heat flow from the baryons to DM as a function of time. The red line indicates the exact solution given by Eqs.~\eqref{eq:heat_conduct_exact_solution} and~\eqref{eq:heat_conduct_kappa}. The simulation results are marked by the dotted lines and the varied parameters are displayed in the legend.}
    \label{fig:convtest_dm_kinetic_sph}
\end{figure}

\begin{figure}
    \centering
    \includegraphics[width=\columnwidth]{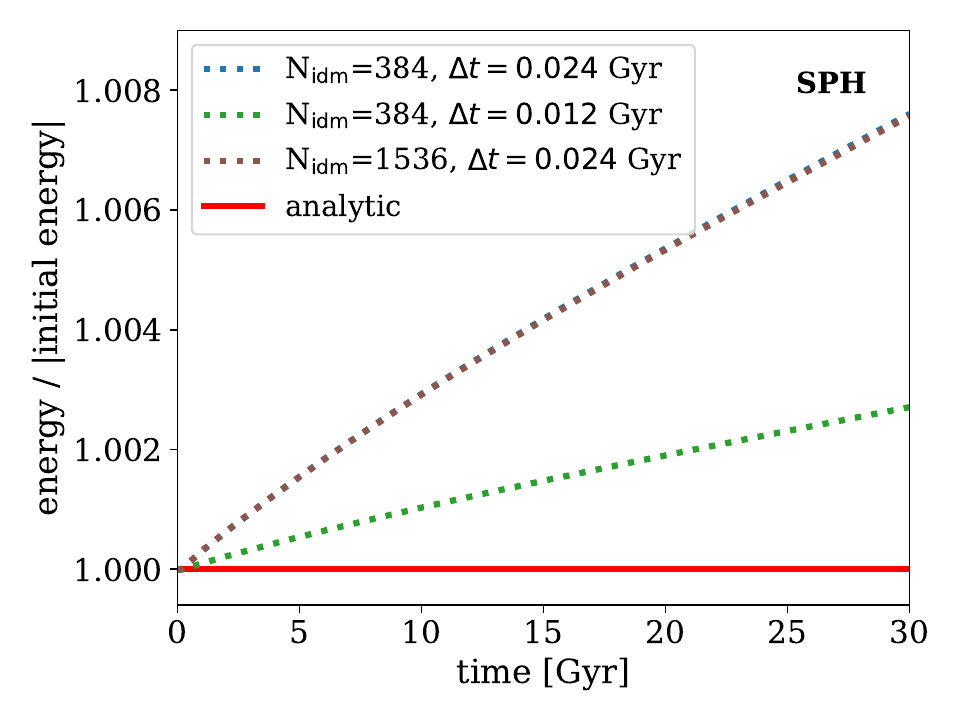}
    \caption{Energy conservation as a function of time of the heat conduction problem for SPH. The results for the same simulations as in Fig.~\ref{fig:convtest_dm_kinetic_sph} are shown. The expectation for perfect energy conservation is indicated by the red line and the numerical results as displayed by the dotted lines with the corresponding simulation parameters being specified by the legend.}
    \label{fig:convtest_energy_conservation_sph}
\end{figure}

\begin{figure}
    \centering
    \includegraphics[width=\columnwidth]{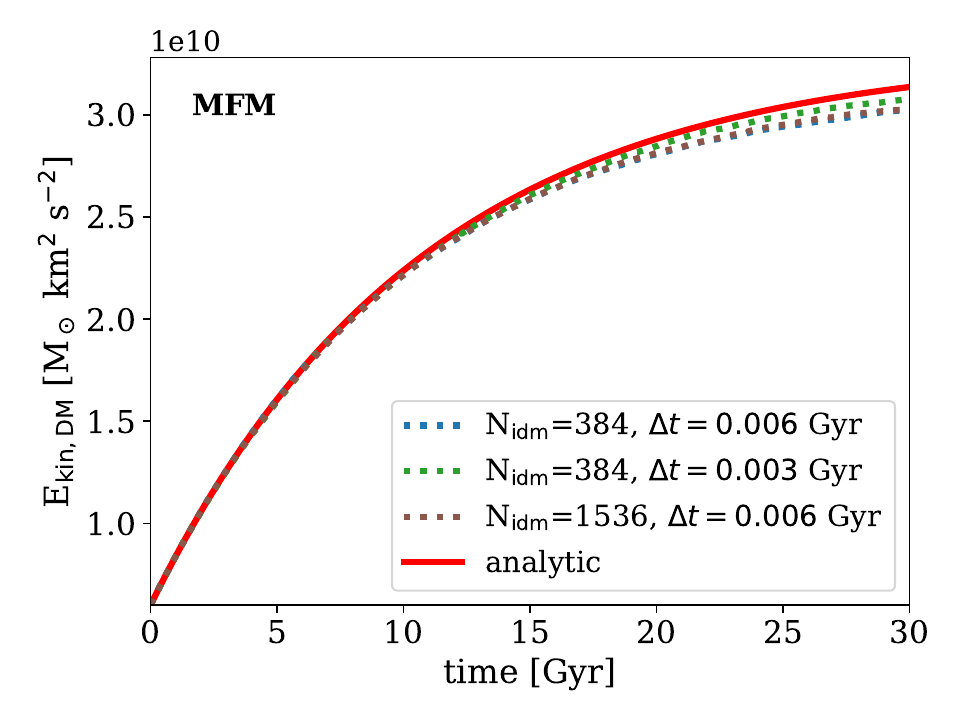}
    \caption{Varying the time step and the interaction number for the heat conduction test problem. The kinetic energy of the DM is shown as a function of time. Simulation results are indicated by the dotted lines and the analytic solution is given by the solid red line. We note this is the same as in Fig.~\ref{fig:convtest_dm_kinetic_sph} but for MFM.}
    \label{fig:convtest_dm_kinetic_mfm}
\end{figure}

\begin{figure}
    \centering
    \includegraphics[width=\columnwidth]{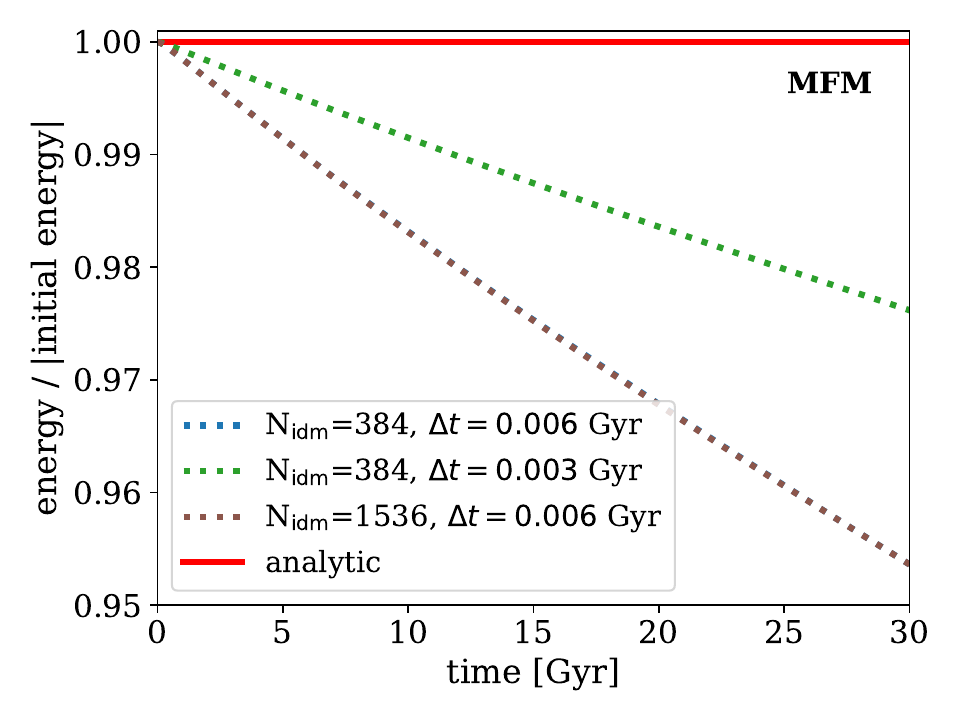}
    \caption{Energy conservation for the heat conduction test problem as a function of time with varying time steps and interaction numbers. The results of the same simulations as used for Fig.~\ref{fig:convtest_dm_kinetic_mfm} are displayed by the dotted lines and the expectation for perfect energy conservation is indicated by the red line.
    We note this is the same as in Fig.~\ref{fig:convtest_energy_conservation_sph} but for MFM.}
    \label{fig:convtext_energy_conservation_mfm}
\end{figure}

In addition to the test shown in Sect.~\ref{sec:test_problems}, we also test if the results of the heat conduction problem with heat flowing from the baryons to the DM changes when varying the interaction number, $N_\mathrm{idm}$, and the size of the time step.

Compared to our fiducial set-up, we run a simulation using half the time step and one using a four times larger interaction number. All SPH simulations were conducted using artificial viscosity and artificial heat conduction, in contrast to the MFM simulations. The set-up is exactly the same as in Sect.~\ref{sec:test_problems_heat_conduct} except for the mentioned parameters.

In Fig.~\ref{fig:convtest_dm_kinetic_sph}, we compare the SPH simulation to the exact solution and find that all runs agree well with the expectation.
Moreover, we also compare the conservation of total energy among the simulations in Fig.~\ref{fig:convtest_energy_conservation_sph}.
Here, we find that increasing $N_\mathrm{idm}$ does only very slightly improve energy conservation. In contrast, decreasing the time step to half its value allows for a significant reduction in energy errors. It is visible that the remaining error is less than half the error in the simulation with the larger time step.
In addition, we also show the results for the MFM simulations in Figs.~\ref{fig:convtest_dm_kinetic_mfm} and~\ref{fig:convtext_energy_conservation_mfm}.

From this, we can conclude that increasing $N_\mathrm{idm}$ beyond 384 does not help much to reduce numerical errors. In contrast, reducing the time step turned out to be more helpful.

\FloatBarrier

\section{Comoving integration test} \label{sec:como_test}

\FloatBarrier

In order to run cosmological simulations we also implemented the comoving integration for the DM-baryon interactions. To test the implementation we simulate a heat conduction problem similar to the tests in Sect.~\ref{sec:test_problems_heat_conduct}, but now in expanding space.

The set-up consists of two homogenous components, DM and baryons, which have the same mass. Together they account for a total mass of $M_\mathrm{tot} = 22.8465 \times 10^{10} \,\mathrm{M_\odot} \, h^{-1}$.
We generate the ICs for a scale factor of $a=0.5$ with the mass residing within a cube with a comoving side length of $1400 \,\mathrm{kpc} \, h^{-1}$, implying a comoving density of $83.26 \, \mathrm{M_\odot} \, \mathrm{kpc}^{-2} \, h^2$. The ICs contain, $N_\mathrm{DM} = 10^5$, numerical DM particle and, $N_\mathrm{bary}=46656$, SPH particles.
The velocities of the DM initially follow a Maxwell--Boltzmann distribution, their kinetic energy sums up to $E_\mathrm{DM} = 2.74969 \times 10^{11} \, \mathrm{M_\odot} \, h^{-1} \, \mathrm{km}^2 \, \mathrm{s}^{-2}$. The bulk velocity of the SPH particle is set to zero but their internal energy makes up $E_\mathrm{bary} = 2.74158 \times 10^{12} \, \mathrm{M_\odot} \, h^{-1} \, \mathrm{km}^2 \, \mathrm{s}^{-2}$.
For the simulation, we switch off gravity but use periodic boundary conditions and an expanding space ($h=0.7$, $\Omega_\mathrm{M,0} = 0.3$, $\Omega_\mathrm{\Lambda} = 0.7$).
Furthermore, we employ artificial viscosity and artificial heat conduction \citep{Price_2012, Beck_2016}.
For the DM-baryon interactions, the default neighbour numbers of, $N_\mathrm{hydro}=230$, $N_\mathrm{ngb} = 64$, and $N_\mathrm{idm} = 384$ are used.
We simulate a cross-section of $\sigma_\mathrm{T}/m_\chi = 2\times 10^6 \,\mathrm{cm}^2 \, \mathrm{g}^{-1}$ employing a forward dominated cross-section (Eq.~\eqref{eq:cross-section_fwd}) with equal masses of the scattering particles ($r=1$).

\begin{figure}
    \centering
    \includegraphics[width=\columnwidth]{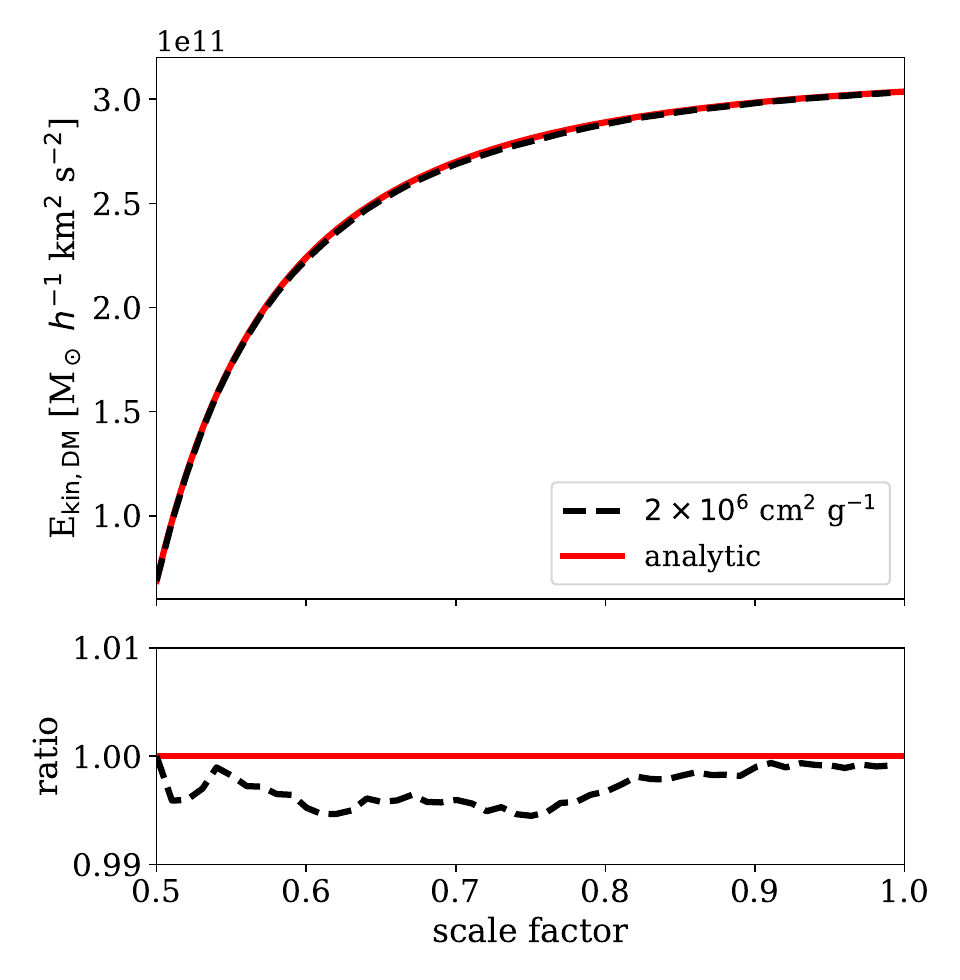}
    \caption{Heat conduction problem for comoving integration. The kinetic energy in terms of the canonical momentum is shown as a function of the scale factor. The simulation results are indicated by the black dashed line and the exact solution (Eq.~\eqref{eq:heat_conduction_cosmo}) is shown in red. The upper panel gives the absolute values, and the lower panel displays the ratios to the exact solution.}
    \label{fig:como_test}
\end{figure}

In Fig.~\ref{fig:como_test}, we show the results for our test set-up. In detail, we display the kinetic energy of the DM as a function of the scale factor, $a$. Here, the kinetic energy is computed as
\begin{equation} \label{eq:kinetic_energy}
    E_\mathrm{kin, DM} = \frac{1}{2} \sum_i m_i \, \mathbf{w}^2 \,,
\end{equation}
with the canonical momentum $\mathbf{w} = a \, \mathbf{v}_\mathrm{pec}$. In contrast to the peculiar velocity, $\mathbf{v}_\mathrm{pec}$, the canonical momentum is conserved while the scale factor is increasing. This implies that also the kinetic energy when expressed by Eq.~\eqref{eq:kinetic_energy} is conserved. As a consequence, all of the change in energy shown in Fig.~\ref{fig:como_test} can be attributed to the DM-baryon interactions and is not affected by the expansion.

To compare the simulation results with the exact solution, we compute the solution building up on the work by \cite{Dvorkin_2014}. In the non-expanding case the power $P_\mathrm{DM}$, arising from the heat conduction is given by
\begin{align} \label{eq:heat_conduction_static}
    P_\mathrm{DM} = \frac{\mathrm{d}E_\mathrm{DM}}{\mathrm{d}t} = -\frac{32}{\sqrt{27\uppi}} \, \frac{\rho_\mathrm{DM} \, \rho_\mathrm{bary}}{(1+r)^2} \frac{\sigma_\mathrm{T}}{m_\chi} \left(\frac{E_\mathrm{DM}}{M_\mathrm{DM}} + \frac{E_\mathrm{bary}}{M_\mathrm{bary}}\right)^{1/2} \nonumber \\
    \times \left[ E_\mathrm{DM} \left( \frac{1}{\rho_\mathrm{DM}} + \frac{r}{\rho_\mathrm{bary}} \right) - \frac{r E_\mathrm{tot}}{\rho_\mathrm{bary}} \right] \,.
\end{align}
For Sect.~\ref{sec:test_problems_heat_conduct}, we computed the full solution analytically (Eqs.~\ref{eq:heat_conduct_exact_solution} and~\ref{eq:heat_conduct_kappa}). For the comoving integration test the differential equation becomes,
\begin{equation} \label{eq:heat_conduction_cosmo}
    \frac{\mathrm{d}E_\mathrm{DM}}{\mathrm{d}a} = \frac{P_\mathrm{DM}(a)}{a \, H(a)} -\frac{2 E_\mathrm{DM}}{a} \,.
\end{equation}
Here we express the energy change as a derivative of the scale factor and we use the Hubble parameter $H(a)$. The first term on the right-hand side of Eq.~\eqref{eq:heat_conduction_cosmo} corresponds to the energy change caused by heat conduction, whereas the second one corresponds to the expansion of space. Moreover, we note that the quantities in Eq.~\eqref{eq:heat_conduction_static} and~\eqref{eq:heat_conduction_cosmo} are in terms of their peculiar values, i.e.\ $\rho \propto a^{-3}$ and $E \propto a^{-2}$.\footnote{We note that this energy definition is different from the one expressed by Eq.~\eqref{eq:kinetic_energy}.}
To obtain the exact solution, we solve the differential equation given by Eq.~\eqref{eq:heat_conduction_cosmo} numerically.

In Fig.~\ref{fig:como_test}, we can see that the simulation result (black) agrees well with the exact solution (red). Hence, we can conclude that our implementation of DM-baryon interactions works for comoving integration.
\end{appendix}

\end{document}